\documentclass[a4paper,twocolumn,11pt,accepted=2023-08-24]{quantumarticle}
\pdfoutput=1
\usepackage[utf8]{inputenc}
\usepackage[T1]{fontenc}
\usepackage{amsmath, amsthm, amssymb, mathtools, dsfont} 
\usepackage{fixmath}
\usepackage{graphicx}  
\usepackage{dcolumn} 
\usepackage{bm,bbm}
\usepackage[normalem]{ulem}
\usepackage[usenames,dvipsnames]{xcolor}
\usepackage{esint}
\usepackage{paralist}
\usepackage[normalem]{ulem}
\usepackage{geometry}
\DeclareGraphicsExtensions{.png,.pdf,.eps}
\graphicspath{ {./figures/}}
\definecolor{myurlcolor}{rgb}{0,0,0.7}
\definecolor{myrefcolor}{rgb}{0.8,0,0}
\usepackage{hyperref}
\usepackage[numbers,sort&compress]{natbib}

\DeclareGraphicsExtensions{.png,.pdf,.eps}
\graphicspath{ {./figures/}} 
\usepackage{afterpage}
\usepackage{epstopdf}
\usepackage{cleveref}
\usepackage[caption=false]{subfig}
\usepackage{ragged2e}
\usepackage{dblfloatfix}
\usepackage{float}

\makeatletter
\let\newfloat\newfloat@ltx
\makeatother

\usepackage{stackengine,scalerel}
\stackMath
\makeatletter

 \theoremstyle{plain}
 
 \theoremstyle{plain}
 \newtheorem{lem}{Lemma}
 \theoremstyle{plain}
 \newtheorem{thm}{Theorem}
 \theoremstyle{plain}
  
 \theoremstyle{plain}
 \newtheorem{exa}{Example}
 \theoremstyle{plain}
 
 \theoremstyle{plain}
 
 \theoremstyle{remark}
 \newtheorem*{rem*}{Remark}
 \theoremstyle{plain}
  \newtheorem{rem}{Remark}

\theoremstyle{plain}
 \newtheorem*{conj*}{Conjecture}
 \theoremstyle{plain}

\newcommand{\ot}{\otimes} 
\renewcommand{\exp}{\mathrm{exp}} 
\DeclareMathOperator{\tr}{tr}   
\newcommand{\HS}{\mathrm{HS}} 

\newcommand{\CC}{\mathbb{C}} 

\renewcommand{\dim}{\text{d}} 

\newcommand{\Herm}{\mathrm{Herm}} 

\newcommand{\pstates}{\mathcal{S}} 

\newcommand{\Uu}{\mathrm{U}} 

\newcommand{\av}{\mathrm{av}} 
\newcommand{\s}{\mathrm{s}} 
\newcommand{\m}{\mathrm{m}} 
\newcommand{\ch}{\mathrm{ch}} 

\newcommand{\I}{\mathbb{I}} 
\renewcommand{\S}{\mathbb{S}} 

\newcommand{\idenC}{\mathcal{I}}

\renewcommand{\H}{\mathcal{H}} 

\renewcommand{\P}{\mathsf{P}} 
\newcommand{\M}{\mathsf{M}} 
\newcommand{\N}{\mathsf{N}} 

\newcommand{\clc}{\mathrm{T}} 

\newcommand{\x}{\mathbf{x}} 

\newcommand{\rbracket}[1]{\left(#1\right)} 
\newcommand{\sbracket}[1]{\left[#1\right]} 

\renewcommand{\H}{\mathcal{H}} 
\newcommand{\T}{\mathcal{T}} 
\newcommand{\J}{\mathcal{J}} 

\newcommand{\iden}{\I}

\newcommand{\p}{\mathrm{\mathbf{p}}} 
\newcommand{\q}{\mathrm{\mathbf{q}}} 

\newcommand{\dtv}{\mathrm{TV}} 
\newcommand{\dtr}{\mathrm{d}_{\mathrm{tr}}} 
\newcommand{\dop}{\mathrm{d}_{\mathrm{op}}} 
\newcommand{\ddiam}{\mathrm{d}_\diamond} 

\newcommand{\expect}[1]{\underset{#1}{\mathbb{E}}} 

\newcommand{\dav}{\mathrm{d}_{\mathrm{av}}} 

\newcommand{\davS}{\dav^{\s}}
\newcommand{\davM}{\dav^{\m}}
\newcommand{\davC}{\dav^{\ch}}

\newcommand{\Psym}[1]{\mathbb{P}_{\mathrm{sym}}^{(#1)}} 

\definecolor{dukeblue}{rgb}{0.0, 0.0, 0.61}
\definecolor{cadmiumgreen}{rgb}{0.0, 0.42, 0.24}

\global\long\global\long\global\long\def\ket#1{\mbox{\ensuremath{|#1\rangle}}}

\global\long\global\long\global\long\def\kb#1#2{\mbox{\ensuremath{\ensuremath{\ensuremath{|#1\rangle\!\langle#2|}}}}}

\makeatother
\renewcommand{\ket}[1]{\left| #1 \right>} 
\newcommand{\ketbra}[2]{\left| #1 \rangle\langle #2 \right|} 

\newcommand\numberthis{\addtocounter{equation}{1}\tag{\theequation}}

\begin{document} 	
\title{Operational Quantum Average-Case Distances}
\author{Filip B. Maciejewski}
\affiliation{Center for Theoretical Physics, Polish Academy of Sciences, Al. Lotnik\'ow 32/46, 02-668
Warszawa, Poland}
\affiliation{Research Institute for Advanced Computer Science (RIACS), USRA, Moffett Field, CA}

\author{Zbigniew Pucha{\l}a}
\affiliation{Institute of Theoretical and Applied Informatics, Polish Academy of Sciences, 44-100 Gliwice, Poland}
\affiliation{Faculty of Physics, Astronomy and Applied Computer Science, Jagiellonian University, 30-348 Krak\'{o}w, Poland}

\author{Micha\l\ Oszmaniec}
\affiliation{Center for Theoretical Physics, Polish Academy of Sciences, Al. Lotnik\'ow 32/46, 02-668
Warszawa, Poland}	
\email{oszmaniec@cft.edu.pl}

\begin{abstract}
We introduce distance measures between quantum states, measurements, and channels based on their statistical distinguishability in generic experiments.
Specifically, we analyze the average Total Variation Distance (TVD) between output statistics of protocols in which quantum objects are intertwined with random circuits and measured in a standard basis.
We show that for circuits forming approximate 4-designs, the average TVDs can be approximated by simple explicit functions of the underlying objects -- the average-case (AC) distances. 
We apply AC distances to analyze the effects of noise in quantum advantage experiments and for efficient discrimination of high-dimensional states and channels without quantum memory. 
We argue that AC distances are better suited for assessing the quality of NISQ devices than common distance measures such as trace distance or the diamond norm.
\end{abstract}

\maketitle

\emph{\textbf{Introduction.}}
 In the era of Noisy Intermediate Scale Quantum (NISQ) devices \cite{Preskill2018}, it is instrumental to have figures of merit that quantify how close two quantum protocols are.
The distance measures commonly used for this purpose, for example, in the context of quantum error correction \cite{DiamondNormQEC2019}, such as trace distance or diamond norm, have an operational interpretation in terms of \emph{optimal statistical distinguishability} between two quantum states, measurements, or channels \cite{Nielsen2010,Distances2005,bengtsson_zyczkowski_2006,Puchala2018optimal}.
While it is natural to consider the optimal protocols when one wishes to distinguish between two objects, alas, in reality, such protocols might be not practical. For example, in general,
they require high-depth, complicated quantum circuits \cite{ComplexityGrowthModels}.
From a complementary perspective, quantum distances are often used to compare an \emph{ideal} implementation (of a state, measurement, or channel) with its \emph{noisy} experimental version.
In this context, using the distances based on optimal distinguishability gives information about the worst-case performance of a device in question. This may be impractical as well -- it is not expected that the performance of typical experiments on a quantum device will be comparable to the worst-case scenario.

In this work, we consider the average Total-Variation (TV) distance between output statistics of two protocols in which random circuits interlace quantum objects of interest 
 (see Figure \ref{fig:diagram_average_case_distances}).  
 This can be thought to mimic the typical circumstances in which quantum states, measurements, or channels appear as parts of quantum-information protocols.
We show that  for a broad class of easy-to-implement random circuits (forming approximate $4$-designs), the average TV distance is approximated by simple explicit functions expressible by degree 2 polynomials in objects in question. 
 We use these functions to define  distance measures between states, measurements, and channels.
The so-defined average-case (AC) distances are thus distance measures that approximate average-case total variation distance.
Contrary to conventional distances such as the trace distance or the diamond norm, the AC distances capture the generic behavior of quantum objects in experiments involving only moderate-depth quantum circuits. This feature can be especially relevant in the context of near-term algorithms, such as the Quantum Approximate Optimization Algorithm (QAOA) \cite{farhi2014qaoa,farhi2019quantum,Harrigan2021QAOA} and Variational Quantum Eigensolver (VQE) \cite{peruzzo2014vqe,Kandala2017VQE,Parrish2019VQE}, as it is expected that generic variational circuits will, on average, have properties of unitary designs \cite{Barren2018}. 
We present numerical results suggesting that AC distances are more suitable for quantifying the impact of imperfections on variational algorithms than the conventional distance measures.

Multiple recent quantum advantage proposals are based on random circuits sampling \cite{Google2019,chinesesupreme2021}.
We apply AC distances to understand the effects of noise on such protocols.
We approach the problem from two sides.
First, the AC distances allow to easily 
\emph{lower} bound the average-case TV distance between the noisy\, distribution\, and\, the\, ideal\, distribution,~thus 
\clearpage
 \onecolumngrid
\begin{@twocolumnfalse}
\vspace{-0.5cm}
\begin{figure*}[!h]
\begin{center}
\captionsetup[subfigure]{format=default,singlelinecheck=on,justification=RaggedRight}
{\includegraphics[width=1.0\textwidth]{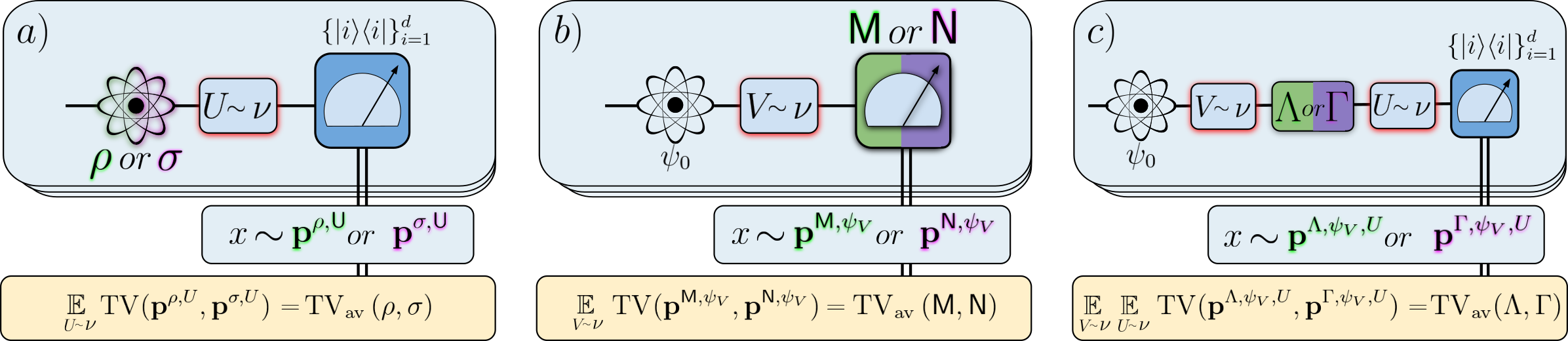}}
\caption{\label{fig:diagram_average_case_distances}
Measures of the distance between quantum objects based on \emph{average} statistical distinguishability.
For quantum states a), we take the average over random unitaries applied to the state, followed by measurement in the standard basis.
For quantum measurements b), we take the average over random pure states measured on the detector.
Finally, for quantum channels c) we take the average over independent random unitaries applied  \emph{before} and \emph{after} the application of the channel.
} \end{center}
\end{figure*}
\end{@twocolumnfalse}
\twocolumngrid 
 \noindent giving insight into how well separated, on average, are noisy distributions from target distributions.
Second, AC distances allow to \emph{upper} bound the average-case TV distance between a noisy distribution and a (trivial) uniform distribution.
This allows to study how quickly the noise makes the average distribution useless.
For example, we show that even in the absence of gate and state-preparation noise, the local, symmetric bitflip error in measurements causes noisy distribution to approach trivial one exponentially quickly in system size.

 Recently there has been a lot of interest in algorithms that use randomized quantum circuits, such as shadow tomography \cite{aaronson2018shadow,Huang2020predicting,hadfield2020shadow,Chen2021shadow,hadfield2021shadow} and randomized-benchmarking \cite{Emerson2005RB,Easwar2010RB,Magesan2012RB,Gambetta2012RB,helsen2019RB,flammia2021ACES}.
  Our results can be employed to quantify the performance of randomized algorithms in the task of statistical distinguishability of quantum objects.
Namely, if the average-case distance between a pair of quantum objects on $N$ qubit systems is large, then they can be (statistically) distinguished almost perfectly using a randomized protocol with just a few implementations of local random circuits of depth $O(N)$. 
We observe that such behavior takes place in two scenarios related to those recently analyzed in the context of so-called Quantum Algorithmic Measurement \cite{QuantumAlghoritmicMeasurements} and complexity growth of quantum circuits \cite{ComplexityGrowthModels}: (i) distinguishing Haar random N qubit pure state from maximally mixed state and (ii) distinguishing N qubit Haar random unitary from maximally depolarizing channel. 
This shows that protocols employing random circuits can be used to efficiently discriminate quantum objects. 
Since they do not depend on the objects to be distinguished, randomized measurement schemes can be interpreted as "universal discriminators", analogous to the SWAP test but not requiring the usage of entanglement or coherent access to copies of quantum systems. 

The manuscript is accompanied by a complementary work \cite{technicalVERSION} that contains proofs of theorems, a thorough analysis of the properties of average-case quantum distances, and further examples.
In contrast, the following work focuses on providing intuition behind AC distances and demonstrating how they can be applied to understand the power of random quantum circuits in practically relevant scenarios, which is followed by numerical demonstrations.

\emph{\textbf{Notation and basic concepts.}}
Our result concern quantum systems on finite-dimensional Hilbert space $\H_d\approx \CC^d$. 
General quantum measurements, also known as POVMs, are described by tuples $\M=(M_i)_{i=1}^n$ of operators on $\H_d$ which satisfy $M_i\geq 0$ and $\sum_{i=1}^n M_i = \iden_d$, where $\iden_d$ is the identity on $\H_d$. 
General quantum operations on $\H_d$ is described by a quantum channel, i.e., a completely-positive trace-preserving map $\Lambda:\Herm(\H_d)\rightarrow \Herm(\H_d)$. We will use the notation  $\tau_d=\iden/d$ to denote maximally mixed state on $\H_d$.

We will consider general protocols consisting of three stages (i) state preparation, in which quantum system is initialized in state $\rho$, (ii) evolution given by a quantum channel $\Lambda$ and (iii) measurement of the resulting state $\Lambda(\rho)$ by a POVM $\M$. 
The outcome statistics of such a protocol are given by the Born rule:  $p_i^{\rho,\Lambda,\M}=\tr(M_i \Lambda(\rho))$.
Total Variation (TV) distance between distributions $\p=(p_i)_{i=1}^n $ and $\q=(q_i)_{i=1}^n$ is defined as $
    \dtv(\p,\q)=\frac{1}{2}\sum_{i=1}^n|p_i-q_i|$. 
TV distance defines the statistical distinguishability of $\p$ and $\q$. Specifically, in a task when we are asked to decide whether the provided samples come from $\mathbf{p}$ or $\mathbf{q}$ (where both are promised to be given with equal probability), the optimal probability of correctly guessing the answer is $p_{\mathrm{succ}}=\frac{1}{2}(1+\dtv(\p,\q))$. 
The related distance between quantum objects is constructed by considering the optimal success probability of distinguishing between pairs of relevant quantum objects, where the optimization is carried out not only over classical post-processing strategies but also over \emph{quantum} strategies that produce classical outcomes given the objects in question (see Supplementary Material (SM) for details). 

Here we propose alternative distance measures based on scenarios where the strategy of discrimination of quantum objects is based on intertwining them with random quantum circuits and then comparing their outcome statistics \cite{technicalVERSION}. Specifically, consider output statistics $\p^{\alpha,\beta}$ of a quantum protocol  
 where $\alpha$ is a fixed quantum object while $\beta$ is taken to be a random variable (specifying a quantum circuit) distributed according to probability distribution $\nu$.
 The average statistical distinguishability  of two objects $\alpha_1,\alpha_2$ is quantified by
\begin{equation}\label{eq:GenAVdefinition}
      \dtv_{\av}(\alpha_1,\alpha_2) = \expect{\beta\sim \nu}\ \dtv(\p^{\alpha_1,\beta},\p^{\alpha_2,\beta})\  .
\end{equation}
Explicit computation of $\dtv_\av(\alpha_1,\alpha_2)$ is difficult because $\dtv(\p,\q)$ is not a polynomial function of the involved probabilities. However, if $\nu$ forms an approximate $4$-design, it is possible to find simple estimates to $\dtv_\av$. Unitary $k$-designs are measures on $\Uu(\H_d)$ that reproduce averages of Haar measure $\mu$ on balanced polynomials of degree $k$ in $U$ \cite{AmbainisEmerson2007}. For approximate $k$-designs these averages agree only approximately. Measure $\nu$ on $\Uu(\H_d)$ is $\delta$-approximate $k$-design if $ \left\| \T_{k,\nu} - \T_{k,\mu} \right\|_\diamond \leq \delta$, where $\mathcal{T}_{k,\nu}(\rho)=\int_{\Uu(\H)} d\nu(U) U^{\ot k} \rho (U^{\dagger})^{\ot k}$. 
Importantly, random quantum circuits in the 1D architecture formed from \emph{arbitrary} universal gates that randomly couple neighboring qubits, generate approximate $k$-designs efficiently with the number of qubits $N$ \cite{LocalRandomCircuitsDesigns,DesignsShallow2018, designsNETS,ExplicitDesignsNickJonas2021}. Specifically, $\delta$-approximate $4$-designs are generated by the 1D random brickwork architecture in depth $O(N+\log(1/\delta))$, with moderate numerical constants \cite{ExplicitDesignsNickJonas2021}.

\emph{\textbf{Quantum average-case distances between states, measurements, and channels.}}
We are now ready to formulate our main
technical results - dimension independent relative error estimates on average $\dtv$ distances between three types of quantum objects depicted in Figure \ref{fig:diagram_average_case_distances}. To simplify the formulation of the Theorems, we will use the symbol $\approx$ to denote equality up to a dimension-independent relative error. 
The specific constants are given in \cite{technicalVERSION}.
In Appendix \ref{app:sec_proofs_theorems} we provide simplified proofs of the following theorems in the setting of exact unitary designs. 
The proofs for approximate unitary designs can be found in Appendix B of \cite{technicalVERSION}.

\emph{Quantum states. }
Let $\p^{\rho,U}$  denote the probability distribution of a quantum process in which $\rho$ undergoes a unitary transformation $U$ and is then subsequently measured in the computational basis of $\H_d$. In other words $p_i^{\rho,U}=\tr\left(\kb{i}{i} U\rho U^\dag\right)$, where $\lbrace\ket{i}\rbrace_{i=1}^d$ is a computational basis of $\H_d$.   

\begin{thm}[Average-case distinguishability of quantum sates -- Theorem~1 from \cite{technicalVERSION}]\label{th:STATESav} Let $\rho,\sigma$ be quantum states in $\H_d$ and let $\nu$ be a distribution in the unitary group $\Uu(\H_d)$ forming $\delta$-approximate $4$-design for $\delta=\frac{\delta'}{2d^4}$, for $\delta'\in(0,\frac{1}{3})$. 
We then have 
\begin{equation}\label{eq:statesBOUNDS}
   \expect{U\sim\nu} \dtv(\p^{\rho,U},\p^{\sigma,U}) \approx \dav^{\s}(\rho,\sigma)=\frac{1}{2}\|\rho-\sigma\|_{\HS}\ ,
\end{equation}
where $\|X\|_{\HS}=\sqrt{\tr(X^2)}$ denotes Hilbert-Schmidt norm. 
\end{thm}

The proof of Theorem \ref{th:STATESav} (and also theorems \ref{th:MEASav} and  \ref{th:CHANNELSav} stated below) is inspired by the proof of  Theorem 4 from \cite{AmbainisEmerson2007} where Berger inequality (stating that for every random variable $X$ with well-defined 2nd and 4th moments we have $ (\expect{} [X^2])^\frac{3}{2}(\expect{} [X^4])^{-\frac{1}{2}}  \leq \expect{} |X|\ $) was used to prove that two states far apart in Hilbert-Schmidt norm can be information-theoretically distinguished by a POVM constructed from approximate $4$-design.

\begin{rem}\label{rem:interpretation}
We can interpret the above average statistical distinguishability as $\dtv$-distance of output statistics resulting from a measurement of a \emph{single} POVM with effects $M_{i,V_j}=\nu_j U^\dagger_j \kb{i}{i} U_j$ , where $\nu_j$ is the probability of occurence of circuit $U_j$ in the ensemble $\nu$ (for simplicity of presentation we assumed that ensemble $\nu$ is discrete). 
This POVM can be interpreted as a convex combination \cite{Oszmaniec17} of projective measurements $\M^{U_j}$ with effects $\M^{U_j}_i=  U^\dagger_j \kb{i}{i} U_j$.
Lower bound on average TV distance implies that such randomized protocol distinguishes between quantum states with high probability.
It immediately follows that there also exists a deterministic (not randomized) \emph{optimal} distinguishability protocol that achieves the same success probability. Such a measurement can be implemented, for example, via Naimark's dilation using an ancillary system \cite{Nielsen2010}.
Analogous interpretation holds also for the average $\dtv$-distances from Theorems \ref{th:MEASav}  and \ref{th:CHANNELSav} below. 
\end{rem}

\begin{rem}
  We note that while the dependence of $\delta$ on the dimension of the system $\dim$ is very high in Theorem~\ref{th:STATESav} (as well as in Theorems~\ref{th:MEASav} and \ref{th:CHANNELSav}), it does not pose a practical problem.
  Indeed, exponentially accurate $\delta$-approximate unitary designs can be implemented already with linear-depth quantum circuits \cite{ExplicitDesignsNickJonas2021}.
\end{rem}

\emph{Quantum measurements. } Let $\p^{\M,\psi_V}$  denote the probability distribution of a quantum process in which a fixed pure quantum state $\psi_0$ is evolved according by unitary $V$ and is subsequently measured via a $n$-outcome POVM $\M=(M_1,M_2,\ldots, M_n)$. In other words $p_i^{\M,\psi_V}=\tr(V\psi_0 V^\dagger M_i)$. 

\begin{thm}[Average-case distinguishability of quantum measurements -- Theorem~2 from \cite{technicalVERSION}]\label{th:MEASav} Let $\M,\N$ be $n$-outcome POVMs on $\H_d$ and let $\nu$ be a distribution on on $\Uu(\H_d)$ forming $\delta$-approximate $4$-design for $\delta=\frac{\delta'}{(2d)^8}$, for $\delta'\in(0,\frac{1}{3})$. We then have 
\begin{align*}\label{eq:povmBOUNDS}
 &\expect{V\sim\nu} \dtv(\p^{\M, \psi_V},\p^{\N,\psi_V}) \approx \dav^\m(\M,\N)\ \ ,\ \text{where} \\ & \dav^\m(\M,\N) =\frac{1}{2d}\sum_{i=1}^n \sqrt{ \|M_i-N_i\|_{\HS}^2 + \tr(M_i-N_i)^2}\  \numberthis . 
\end{align*}
\end{thm}

\emph{Quantum channels.}
Let $\p^{\Lambda,\psi_V,U}$ by the probability distribution associated to a quantum process in in which a fixed pure quantum state $\psi_0$ is subsequently acted on by unitary $V$,  channel $\Lambda$  and unitary $U$, and is subsequently measured in the computational basis of $\H$. In other words we have $p_i^{\Lambda,\psi_V,U}=\tr(\kb{i}{i} U \Lambda(V\psi_0 V^\dagger ) U^\dagger)$.

\begin{thm}[Average-case distinguishability of quantum channels -- Theorem~3 from \cite{technicalVERSION}]\label{th:CHANNELSav} Let $\Lambda,\Gamma$ be quantum channels acting on $\H_d$. let $\nu$ be a distribution on on $\Uu(\H_d)$ forming $\delta$-approximate $4$-design for $\delta=\frac{\delta'}{(2d)^8}$, for $\delta'\in(0,\frac{1}{9})$.  Then we have 
\begin{align*}\label{eq:channelsBOUNDS}
  &\expect{V\sim\nu} \expect{U\sim\nu} \dtv(\p^{\Lambda,\psi_V,U},\p^{\Gamma,\psi_V,U}) \approx  \dav^{\ch}(\Lambda,\Gamma)\ \ ,\  \text{where}\  \\ &  \dav^{\ch}(\Lambda,\Gamma) = \frac{1}{2} \sqrt{\left\| \J_\Lambda - \J_\Gamma \right\|_{\HS}^2+\tr\left((\Lambda-\Gamma) [\tau_d]^2 \right)\  \numberthis } 
\end{align*}
and $\J_\Lambda$ denotes Jamio\l{}kowski-Choi state of $\Lambda$.
\end{thm}

\begin{rem}
Having defined randomized distinguishability strategies, it is natural to ask how they 
compare to optimal protocols on a $d$-dimensional Hilbert space $\H_d$.
We give upper bounds on the maximal ratio between worst-case and average-case distances
to answer this. 
It turns out that this ratio is at most $d^{\frac{1}{2}},\ d, \ d^{\frac{3}{2}}$ for 
quantum states, measurements, and channels, respectively.
This implies that there exist scenarios where the optimal protocol for distinguishing two quantum objects performs exponentially better than protocol using random quantum circuits.
Indeed, in the technical version of the manuscript, \cite{technicalVERSION} we construct examples
that saturate those bounds.
\end{rem}

The above theorems suggest to define average-case distances between quantum states, measurements, and channels via formulas $\dav^\s,\ \dav^\m,\ \dav^\ch$ appearing in approximations \eqref{eq:statesBOUNDS}, \eqref{eq:povmBOUNDS}, and \eqref{eq:channelsBOUNDS}. 
This approach has several pleasant consequences. 
First, functions describing these distances can be expressed via simple, degree-two polynomials in underlying objects and can be easily explicitly computed for objects acting on systems of moderate dimension (no optimization is needed as in the case of the diamond norm \cite{watrous2009semidefinite}). 
Second, all average-case distances utilize in some way the Hilbert-Schmidt norm. 
This gives this norm an operational interpretation it did not possess before (especially for quantum states for which $\dav^\s(\rho,\sigma)=\frac{1}{2}\|\rho-\sigma\|_\HS$). 
Third, it turns out that so-defined distances satisfy plethora of natural properties such as subadditivity: $\dav^\s(\rho_1\otimes \rho_2 ,\sigma_1\otimes \sigma_2)~\leq~\dav^\s(\rho_1,\sigma_1)+\dav^\s(\rho_2,\sigma_2)$, joint convexity: $\dav^\s(\sum_\alpha p_\alpha \rho_\alpha, \sum_\alpha p_\alpha \sigma_\alpha)\leq \sum_\alpha p_\alpha \dav^\s(\rho_\alpha,\sigma_\alpha)$, or restricted data-processing inequalities (typically various distances $\dav$ are non-increasing under application of unital quantum channels).
See 
\cite{technicalVERSION} for details and proofs of various properties of average-case distances.
Fourth, while it may seem that condition of being (approximate) $4$-design is quite stringent, from a recent paper \cite{ExplicitDesignsNickJonas2021} it follows that ensembles of quantum circuits required by Theorems \ref{th:STATESav}-\ref{th:CHANNELSav} can be realized by random circuits in the 1D brickwork architecture in depth $O(N)$ (with moderate prefactors) \cite{ExplicitDesignsNickJonas2021}. 
Finally, we expect that our average-case distances will more accurately capture the behavior of errors in the performance of quantum objects in generic moderate size quantum algorithms (note that many architectures of variational circuits used in NISQ algorithms are expected to exhibit, on average, design-like behavior \cite{Barren2018}). 
We back up this last claim numerically by testing the usefulness of our distance measures on families of random quantum circuits originating from random instances of variational quantum algorithms on few-qubit systems.

\emph{\textbf{Applications.}} 
For all the reasons mentioned above, we believe that introduced distances will prove useful in analyzing the practical performance of near-term quantum processors. 
We expect that they can also be useful in other branches of quantum information requiring the usage of randomized protocols like quantum communication, quantum complexity theory, or quantum machine learning. 
The following simple examples illustrate potential usefulness of our results.

\textit{Application 1: Noise in quantum advantage experiments.}

Here we consider examples which help to understand how noise affects average probability distributions in experiments with random circuits sampling.
First, AC distances between noisy and ideal state allow to lower-bound average TVDs between target and noisy distributions.
Second, AC distances allow to upper-bound average-case TVD between noisy distribution and trivial (uniform) one.
Indeed, to bound average TVD between uniform and noisy distribution, one calculates AC distance to maximally mixed state $\frac{\iden}{d}$ (states), trivial POVM $\M^{\idenC}=\rbracket{\frac{\iden}{d},\dots,\frac{\iden}{d}}$ (measurements), or maximally depolarizing channel $\Lambda_{\mathrm{dep}}$ that acts as $\Lambda_{\mathrm{dep}}(\rho) = \frac{\iden}{d}$ for any state $\rho$ (channels). 
This follows directly from definitions of AC distances -- see Lemmas~23, 24 and 25 in  \cite{technicalVERSION}.

In what follows, most of the examples make use of some average noise parameter $q^{av}$ (with different meaning for each example) that describes an average (over qubits) probability of errors of considered type \emph{not} occurring.
In most of them, we make an assumption that $q^{av}\leq \sqrt[N]{\frac{1}{2}}$.
This is done solely to achieve a particularly appealing form of lower bounds.
One can derive expressions that are more complicated and do not require this assumption (see SM for details and proofs of the following examples).
In general, since $\sqrt[N]{\frac{1}{2}}\xrightarrow{N\rightarrow \infty}1$, the assumption becomes less restrictive for higher-dimensional systems and the presented bounds are intended for use in such cases.

\begin{exa}[Pauli eigenstates and tensor product Pauli noise]\label{ex:uniform_separable_states}
Consider state $\psi^{\text{pauli}} = \otimes_{i=1}^{N} \ketbra{\pm r_i}{\pm r_i}$, where $r_i \in \left\{x,y,z\right\}$, i.e., $\ket{\pm r_i}$ is any Pauli eigenstate on qubit $i$ (with eigenvalue $+1$ or $-1$.).
Consider tensor product Pauli channel $\Lambda^{\text{pauli}}=\otimes_{i=1}^N \Lambda_{i}^{\text{pauli}}$, where single-qubit channel is $\Lambda^{\text{pauli}}_{i}(\rho) = \sum_{j=1}\ p^{(i)}_j\sigma_j\rho\sigma_j$ with $j \in \left\{1,x,y,z\right\}$, $\sigma_1=\iden$, and $p^{(i)}_j\geq 0$, $\sum_{j}p^{(i)}_j=1$.
Define $q^{(i)} = p^{(i)}_1+p^{(i)}_{r_i}$, i.e., a probability of applying on qubit $i$ a gate that stabilizes the state of that qubit (namely, either identity or Pauli matrix of which $\ket{\pm r_i}$ is an eigenstate).
Define average properties of noise as $q^{av} = \frac{1}{N}\sum_{i=1}^{N}q^{(i)}$ and $f^{av}=\frac{1}{N}\sum_{i=1}^N q^{(i)}(1-q^{(i)})$.
Assume $q^{(i)}\geq\frac{1}{2}$ for each qubit and that $q^{av}\leq \sqrt[N]{\frac{1}{2}}$.
Then we have
\begin{align}\label{eq:states_uniform_example}
    \davS(\Lambda^{\text{pauli}}(\psi^{\text{pauli}}), \frac{\iden}{d}) < \frac{1}{2}\ \exp\left(-2f^{av}\ N\right),
\end{align}
\vspace{-0.75cm}
\begin{align*}\label{eq:states_pauli_example}
        &\davS(\Lambda^{\text{pauli}}(\psi^{\text{pauli}}), \psi^{\text{pauli}})> 
         \frac{1}{2}\sqrt{1-2(q^{av})^N} , \numberthis
\end{align*}
\end{exa}

The above example might be relevant, for example, in QAOA algorithms where input state is often indeed a tensor product Pauli state \cite{farhi2014qaoa}, or can be useful for estimating effects of state-preparation errors for standard setting where input state is $\ketbra{0}{0}^{\otimes N}$.
We see that with growing system size, the average noisy distribution approaches uniform distribution exponentially quickly (while moving away from target distribution).

This demonstrates that even in the absence of noise in random unitaries, the state-preparation errors will quickly aggregate.
Exactly the same behaviour is demonstrated for the following simplified measurement noise model.
\begin{exa}[Symmetric bitflip measurement noise]\label{ex:uniform_separable_povms}
Consider a noisy version $\clc^{\text{sym}} \P$ of computational basis measurement $\P$, where $\clc^{\text{sym}} = \otimes_{i=1}^{N}\clc_i^{\text{sym}}$
and $k$th effect of noisy measurement is given by $\left(\clc^{\text{sym}}\P\right)_k=\sum_{l}\clc^{\text{sym}}_{kl}\ketbra{l}{l}$.
Here for each qubit we have $\clc^{\text{sym}}_i = p^{(i)}\iden+(1-p^{(i)}) \sigma_x$, where $(1-p^{(i)})$ is a bitflip error probability on $i$th qubit.
Define $f^{av}=\frac{1}{N}\sum_{i=1}^{N}p^{(i)}(1-p^{(i)})$.
Assume $p^{(i)}\geq \frac{1}{2}$ for each qubit.
Then we have
\begin{align}\label{eq:uniform_povms_symmetric}
    \davM(\clc^{\text{sym}} \P, \M^{\idenC}) < \frac{1}{2}\exp\left(-2f^{av}\ N\right)\ , 
    \end{align}
\end{exa}
The above means that even in the absence of state-preparation and gate errors, for symmetric bitflip noise the resulting average distribution exponentially quickly converges to uniform.
We now consider a distance from ideal measurement for more realistic case of generic tensor product measurement noise.

\begin{exa}[Generic tensor product measurement noise]\label{ex:meas}
Let $\P=(\kb{\x}{\x})_{\x\in\lbrace0,1\rbrace^N}$ be a computational basis measurement on $N$ qubit system. Let $\M=(M_\x)_{\x\in\lbrace0,1\rbrace^N}$ be a POVM specified by effects $M_\x = \Lambda_1^\dagger(\kb{x_1}{x_1})\otimes \ldots \ot  \Lambda_N^\dagger(\kb{x_N}{x_N})$, where $\Lambda_i$ are quantum channels affecting $i$'th qubit, and $\Lambda^\dagger_i$ is the conjugate of $\Lambda_i$. 
Define classical success probability as $p^{(i)}(k|k) = \tr{\left(\Lambda_{i}^{\dag}\left(\ketbra{x_i}{x_i}\right) \ketbra{x_i}{x_i}\right)}$ and corresponding average  $q_{av}^{(i)} = \frac{p^{i}(0|0)+p^{(i)}(1|1)}{2}$.
Let $q^{av} \coloneqq \frac{1}{N}\sum_{i=1}^{N} q_{av}^{(i)} $.
Assume that for each qubit $q_{av}^{(i)}\geq \frac{1}{2}$ and that $q^{av}\leq \sqrt[N]{\frac{1}{2}}$.
Then we have 
\begin{align}\label{eq:generic_measurement_lower_bound}
    \davM(\M,\P) > \frac{1}{2}\sqrt{1-2(q^{av})^N} \ .
\end{align}
\end{exa}
The quantity $q^{av}$ is the survival probability of classical single-qubit state
$\kb{x_i}{x_i}$  that goes through a channel $\Lambda_i$, averaged over all qubits and input states. 
We note that those quantities are routinely reported in experimental works, which makes the above bound particularly useful. 
Indeed, data from recent quantum advantage experiments \cite{Google2019,chinesesupreme2021}
suggests that $q^{av}$ is around $97\%$ (we take average of values reported in both papers). 
Assume perfect gates, no state preparation errors and $q^{av}=0.97$. Furthermore, assume that random circuits used in experiments form approximate $4$-designs (this assumption is consistent with results of \cite{DesignsShallow2018}). 
Then from Theorem~\ref{th:MEASav} it follows that if
readout errors remain constant with scaling of the system, for a 54-qubit quantum computer, on average (over realizations of random quantum circuits) output distributions $\p^{\M,\psi_V}$ will have a constant $\approx 0.13$ $\dtv$-distance from the ideal probability distributions $\p^{\P,\psi_V}$ solely due to effects of readout noise.

\begin{exa}[Tensor product Pauli noise in the middle of the circuit]\label{ex:uniform_separable_channels}
Consider tensor product Pauli channel $\Lambda^{\text{pauli}}$ defined in Example~\ref{ex:uniform_separable_states}.
For each qubit $i$ define
$||\mathbf{p}^{(i)}||^2_2 = \sum_{j}\left(p^{(i)}_j\right)^2$, and corresponding average $p_2^{av} = \frac{1}{N}\sum_{i=1}^{N}||\mathbf{p}^{(i)}||_2^2 $, as well as average probability of application of identity  channel $p_1^{av} = \frac{1}{N}\sum_{i=1}p^{(i)}_1$.
Assume $p_1^{av}\leq \sqrt[N]{\frac{1}{2}}$.
Then we have
\begin{align}\label{eq:channels_uniform_example}
    \davC(\Lambda^{\text{pauli}}, \Lambda_{\text{dep}}) & < \frac{1}{2}\exp\left(-p_2^{av}\ N\right)
    \ , \\
    \davC(\Lambda^{\text{pauli}}, \idenC) & 
    > \frac{1}{\sqrt{2}}\sqrt{1- 2\left(p_1^{av}\right)^N} \ .
\end{align}
\end{exa}
Recall that the above scenario corresponds to inserting local Pauli noise "between" two random circuits (two averages in Eq.~\eqref{eq:channelsBOUNDS}).
Similarly to previous cases, whenever there is non-zero noise, we will observe an exponential convergence to the trivial distribution and high separation from ideal distribution corresponding to identity channel $\idenC$.

\begin{exa}[Single Pauli error the middle of the circuit]\label{ex:single_qubit_pauli}
Consider tensor product channel $\Lambda^{(i)}_{\sigma}$ that applies some traceless unitary $\sigma$ on qubit $i$ (and identity to all other qubits).
Then we have
\begin{align}\label{eq:states_pauli_single_error}
    \davC(\Lambda^{(i)}_{\sigma}, \idenC) = \frac{1}{\sqrt{2}} \ .
\end{align}
\end{exa}
Physically, the above may correspond to a unitary noise applying one of Pauli matrices on qubit $i$ somewhere in the circuit.
We then observe a constant separation (value of $\frac{1}{\sqrt{2}}$) 
between ideal distribution and the noisy distribution.
Such significant average distance between noisy and target distribution suggests that local strong coherent errors can dramatically affect the performance of a given device in typical circumstances.
This result is in agreement with empirical observations made in Refs.~\cite{Boixo2018,Google2019} where single-qubit errors were causing "speckle pattern" of output bitstrings probabilities to break, resulting in very low cross-entropy benchmarking fidelity.

\textit{Application 2: Sample efficient distinguishability of quantum objects with incoherent access}

\begin{exa}
 \label{ex:stateMIXEDstate}
For any pure state $\psi$ on $\H_d$ we have $ \dav^\s\left(\psi,\tau_d\right) = \frac{1}{2}\sqrt{1-\frac{1}{d}}$. 
\end{exa}

It follows that a single round of a randomized protocol implicit in the definition of $\dav^\s$ (cf. Remark \ref{rem:interpretation}), realized via  approximate $4$-design and computational basis measurements, gives a constant bias in distinguishing \emph{any} pure $N$ qubit state $\psi$ from the maximally mixed state: $p^{\mathrm{av}}_\mathrm{succ}\gtrsim 0.57$. This probability can be made arbitrarily close  $1$ by repeating the protocol and using the majority-vote strategy.  
Importantly, this method \emph{does not} utilize the coherent access or a quantum memory (in a sense defined, e. g., in \cite{QuantumAlghoritmicMeasurements,QuantumAdvML2021}).
We note that a related but distinct scenario is considered in Ref.~\cite{QuantumAlghoritmicMeasurements}. 
There, the authors introduced the task of \texttt{PurityTesting}
corresponding to discrimination
between \emph{unknown} Haar-random pure random state and maximally mixed state.
For $N$ qubit systems, Theorem 4 of \cite{QuantumAlghoritmicMeasurements} implies exponential lower bound for the query complexity $k$ (number of usages of unknown quantum state) needed to succeed in this task, given incoherent access to objects in question. 
In contrast, our randomized measurement protocol gives high statistical distinguishability already for a single query \emph{for all} states $\psi$. 
The difference comes from the fact that in the scenario considered in Example~\ref{ex:stateMIXEDstate} the random state is arbitrary but known.

\begin{exa}\label{ex:unitaryDEPL}
Let $\Lambda_U$ be a a unitary channel corresponding to a unitary $U$ on $\H_d$ and let $\Lambda_\mathrm{dep}$ be a depolarizing channel i.e. $\Lambda_{\mathrm{dep}}(\rho)=\tau_d$ for any $\rho$. Then we have  $\dav^\ch\left(\Lambda_U,\Lambda_{\mathrm{dep}}\right)  = \frac{1}{2}\sqrt{1-\frac{1}{d^2}}$.
\end{exa}
In related task \texttt{FixedUnitary} studied in \cite{QuantumAlghoritmicMeasurements}, one is asked to distinguish \emph{unknown} Haar-random unitary channel $\Lambda_U$ from $\Lambda_{\mathrm{dep}}$.
Exponential query complexity lower bound incoherent protocols was shown in  \cite{QuantumAlghoritmicMeasurements}.
By repeating analogous reasoning as for states, we get that when 
$\Lambda_{U}$ is arbitrary but known, randomized, non-adaptive, and incoherent protocol, utilizing two realizations of approximate $4$-designs, gives constant bias in 
success probability of discrimination of $\Lambda_U$ from $\Lambda_{\mathrm{dep}}$ using just a single query. 

\textit{Application 3: Strong complexity of quantum states and unitaries. }
The above o examples have interesting consequences for the notion of a strong state and unitary complexity investigated in \cite{ComplexityGrowthModels}. 
There, the authors defined complexity $C_\Delta$ of $N$-qubit pure state $\psi$ (resp. unitary circuit $\Lambda_U$) as the number of elementary gates needed to construct a circuit necessary to implement a \emph{two-outcome} measurement discriminating between $\psi$ (resp. depolarizing channel $\Lambda_{\mathrm{dep}}$) with success probability $p_{\mathrm{succ}}=\frac{1}{2}+\Delta$. Our results imply that if the requirement of two-outcome measurement is relaxed, then measurements\, realizable\,\, with\,\, circuit\, depths\,\, $r=\mathrm{poly}(N)$ 
\clearpage
\onecolumngrid
\begin{@twocolumnfalse}
\begin{figure*}[!t]
\begin{center}
\captionsetup[subfigure]{format=default,singlelinecheck=on,justification=RaggedRight}
\subfloat[\label{fig:numerics_scaling_states_ideal}Quantum states, distance to ideal distribution]
        {\includegraphics[width=0.475\textwidth]{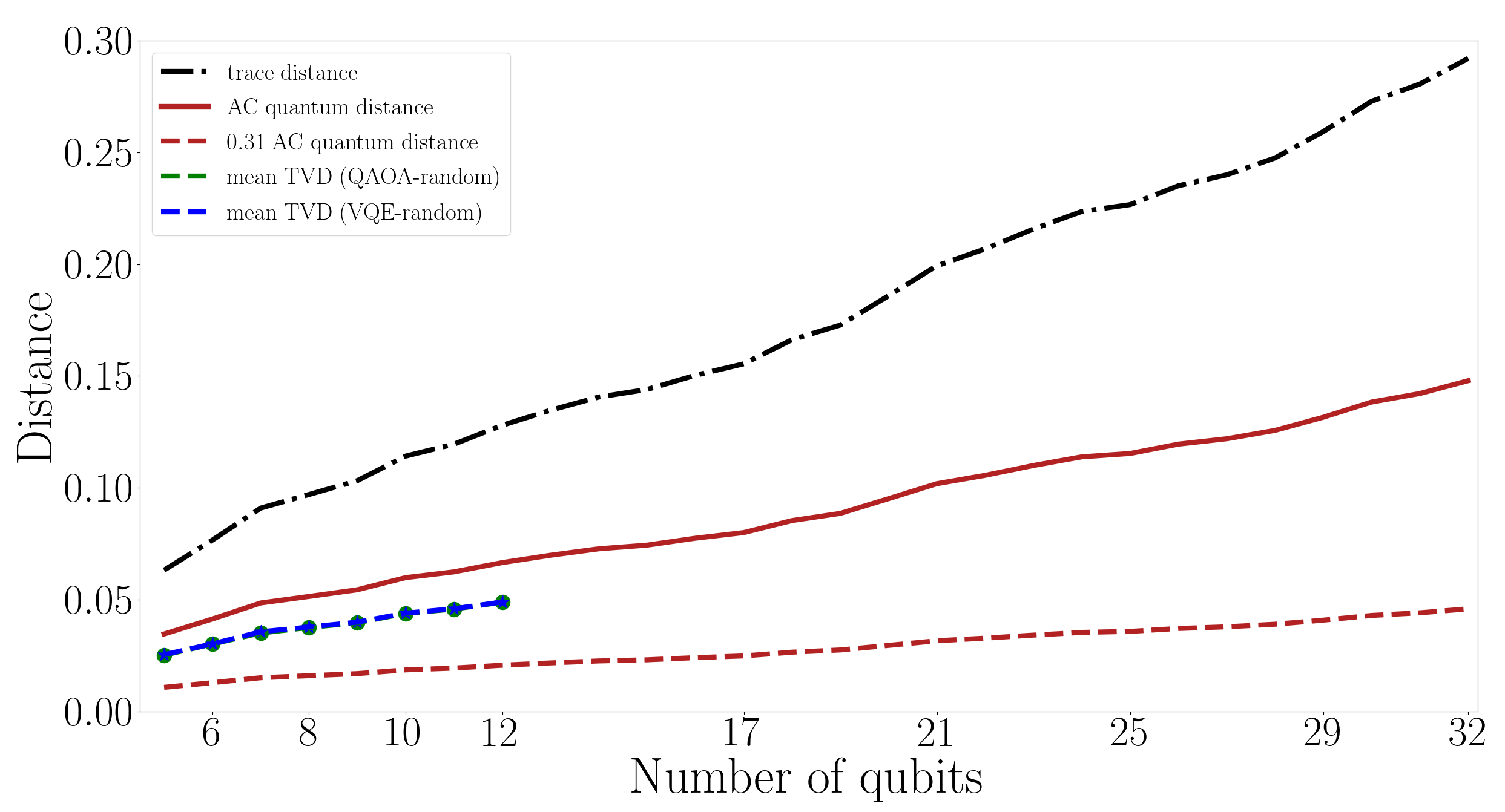}}
\subfloat[\label{fig:numerics_scaling_states_uniform}Quantum states, distance to uniform distribution]
        {\includegraphics[width=0.475\textwidth]{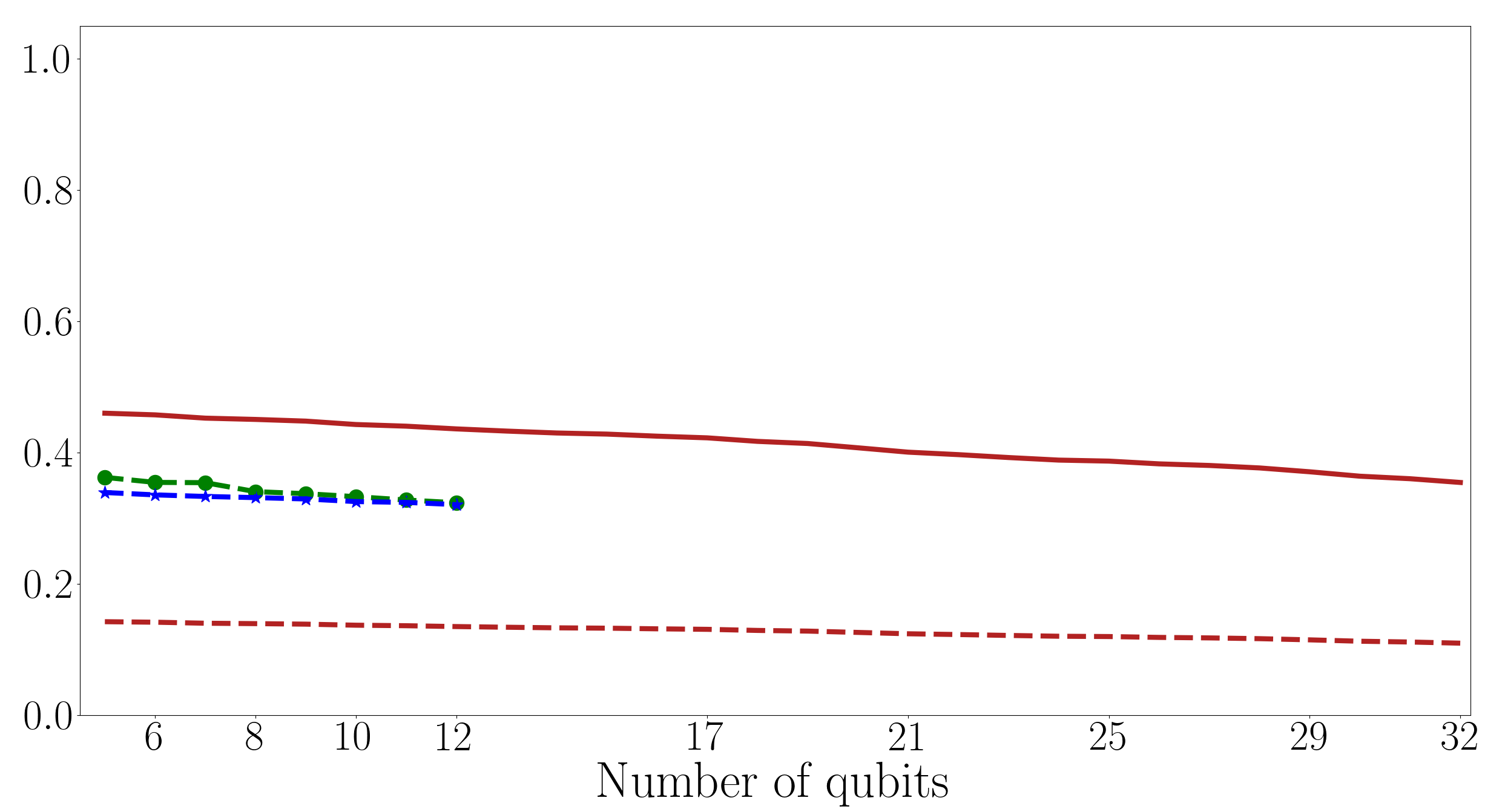}}
        \\
\subfloat[\label{fig:numerics_scaling_measurements_ideal}Quantum measurements, distance to ideal distribution]
    {\includegraphics[width=0.475\textwidth]{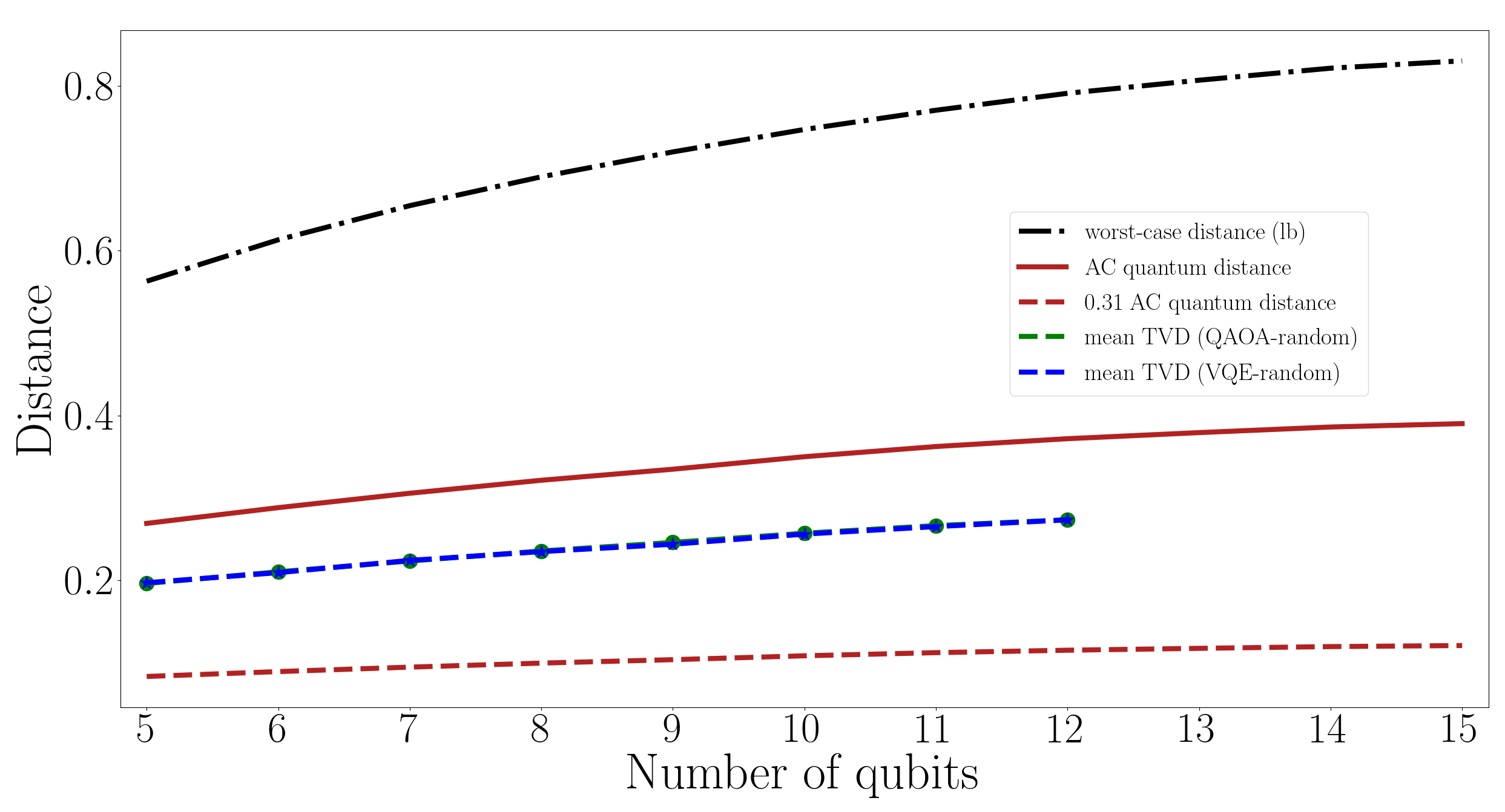}}
\subfloat[\label{fig:numerics_scaling_channels_ideal}Quantum channels, distance to ideal distribution]
        {\includegraphics[width=0.475\textwidth]{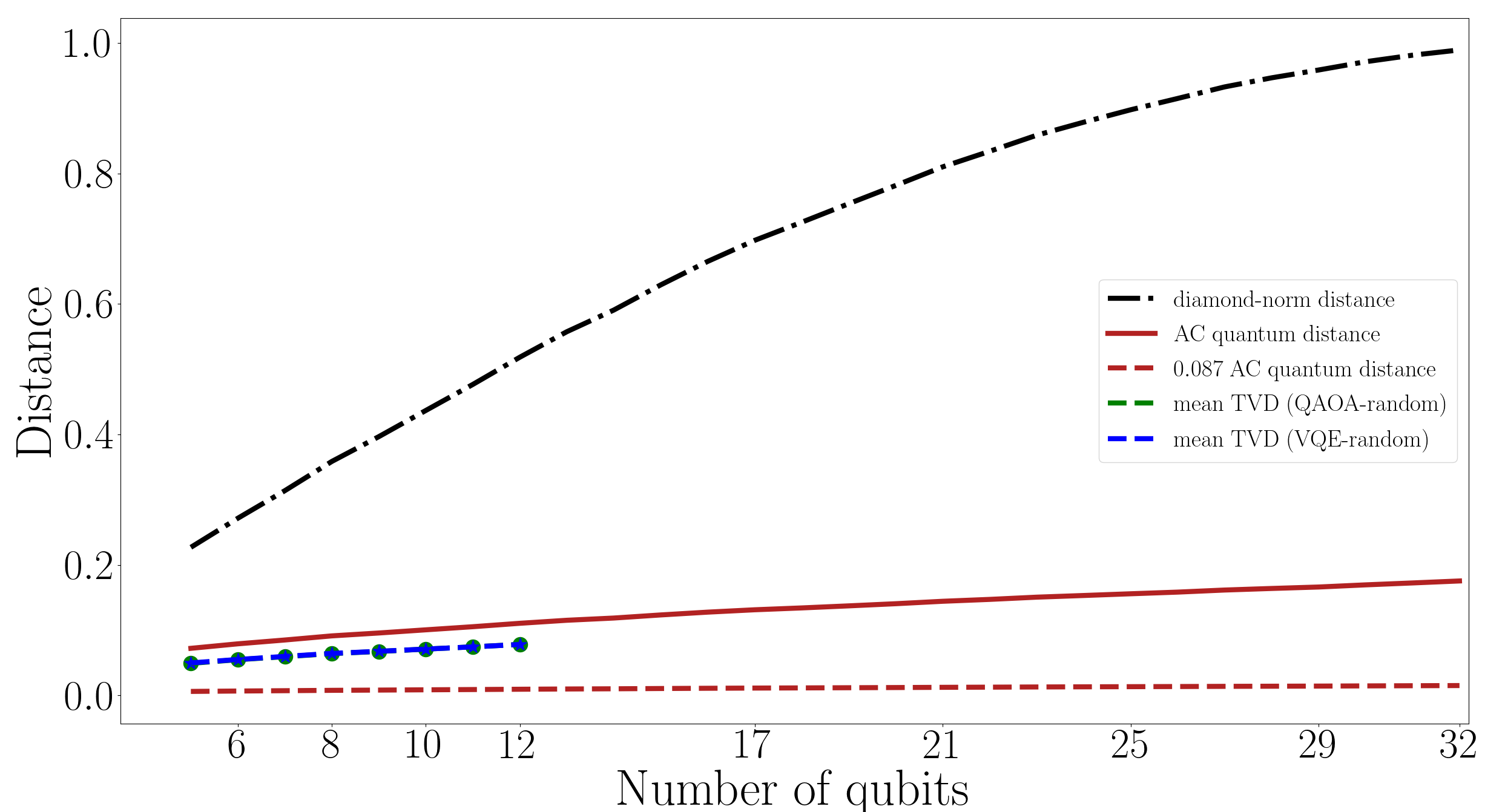}}
    \caption{\label{fig:numerics_scaling}
Results of numerical studies for comparison between worst-case distance, average-case quantum distance and numerically calculated mean TVD.
Plots ~\ref{fig:numerics_scaling_states_ideal},~\ref{fig:numerics_scaling_measurements_ideal} and~\ref{fig:numerics_scaling_channels_ideal} correspond to distance to ideal (noiseless) distribution.
For states, we additionally plot distance to uniform (trivial) distribution on plot~\ref{fig:numerics_scaling_states_uniform}.
For average-case distance, we also plot value corresponding to lower bound on average-case TVD (following from Eqs.~\eqref{eq:statesBOUNDS}, \eqref{eq:povmBOUNDS}, \eqref{eq:channelsBOUNDS}).
In case of worst-case distance, "lb" indicates lower-bound.
Average-case quantum distances were calculated explicitly.
Mean TVDs were calculated between (exact numerical) probability distributions over $1000$ random instances of random unitaries. 
}
 \end{center}
\end{figure*}
\end{@twocolumnfalse}
\twocolumngrid
\noindent can succeed in these discrimination tasks with a constant bias $\Delta_\ast$ \emph{for all} states $\psi$ and unitary channels $\Lambda_U$.
This renders the so-defined notion of complexity trivial - all states and unitaries will have complexity $C_\Delta\leq\mathrm{poly}(N)$, unless bias $\delta$ satisfies  $\Delta>\Delta_\ast$.  

We note that large average-case distance  $\dav$ implies only information-theoretic distinguishability of quantum objects. 
The cost of classical post-processing needed to distinguish the probability distributions resulting from randomized protocols can be very large since they operate on exponentially large sample space.

\emph{\textbf{Numerical results.}}
Here we present the results of numerical studies of small-size quantum systems.
We compare scaling with the system size for worst-case distance, average-case distance, and a mean TVD taken over an ensemble of random unitaries.
The mean Total-Variation distance is calculated numerically over two types of ensembles of unitaries with a structure of variational circuits.
One ensemble has a QAOA-like structure, while the other is a standard hardware-efficient VQE ansatz \cite{peruzzo2014vqe}, both initialized with random parameters (see SM for exact form).
Based on recent results \cite{Barren2018}, we expect them to form (approximate) unitary $4$-designs.

We consider the following scenarios. 
\begin{enumerate}
    \item (States) 
We compare a randomly chosen Pauli eigenstate affected by random local Pauli noise with its ideal version (Fig.~\ref{fig:numerics_scaling_states_ideal}) and with maximally mixed state $\frac{\iden}{d}$ (Fig.~\ref{fig:numerics_scaling_states_uniform} ).
This is the scenario considered in Example~\ref{ex:uniform_separable_states}.
The error probabilities are chosen randomly from range $\sbracket{0.001,0.01}$.
 \item (Measurements)
The noisy measurement is a tensor product POVM constructed from single-qubit measurements obtained via Quantum Detector Tomography \cite{Lundeen2008} of IBM's 15-qubit Melbourne device.
We compare it to ideal computational-basis measurement (Fig.~\ref{fig:numerics_scaling_measurements_ideal}).
Since the measurement noise in superconducting devices is usually highly asymmetric \cite{Maciejewski2021}, we do not expect it to converge to the uniform distribution.
\item 
(Channels) 
We compare channel corresponding to random tensor product 1-qubit rotations around a random axis with ideal identity channel $\idenC$ (Fig~\ref{fig:numerics_scaling_channels_ideal}).
Explicitly, the unitary corresponding to the channel has a form $\bigotimes_{k=1}^{N}\exp(-i \gamma_k V^{(k)})$, where $V^{(k)}$ is chosen randomly to be $X$, $Y$ or $Z$ gate, and  $\gamma_{k} \in \sbracket{0.025\pi, 0.0313\pi }$.
Similarly to POVMs, we do not expect coherent errors to bring noisy distributions close to the uniform distribution.
\end{enumerate}
In each case, the number of circuit layers is  $\lfloor1.5N\rfloor$.
In Fig.~\ref{fig:numerics_scaling} we collectively present the results of all simulations.
Recall that both ensembles presented in Fig.~\ref{fig:numerics_scaling} consist of circuits that are variational QAOA and VQE circuits with random parameters.
From the plots, it is clear that in all studied cases for those ensembles, the average-case quantum distance is both significantly closer and more similar in scaling to the mean Total Variation distance between distributions in question, as compared
to worst-case distance.

\begin{acknowledgments}
\emph{Acknowledgements} 
We would like to thank Richard Kueng, Victor Albert, and Ingo Roth for interesting discussions and comments.  We are grateful to Jordan Cotler for clarifying tasks of \texttt{PurityTesting} and \texttt{FixedUnitary} considered in Ref.~\cite{QuantumAlghoritmicMeasurements}. 
We sincerely thank Susane Calegari for proofreading the manuscript.
We used Qiskit \cite{qiskit} and QuTiP \cite{qutip1,qutip2} to perform some of the simulations.

The authors acknowledge the financial support by the TEAM-NET project co-financed by the EU within the Smart Growth Operational Programme (contract no. POIR.04.04.00-00-17C1/18-00). 
We are grateful to Jordan Cotler for improving our understanding of Purity Testing and Fixed Unitary tasks considered in Ref.~\cite{QuantumAlghoritmicMeasurements}. 
We used Qiskit \cite{qiskit} and QuTiP \cite{qutip2} to perform some of the simulations.
\end{acknowledgments}

\bibliographystyle{plainurl}
\bibliography{nisqDISTbib.bib} 

\begin{thebibliography}{10}

\bibitem{aaronson2018shadow}
Scott Aaronson.
\newblock Shadow tomography of quantum states.
\newblock {\em SIAM Journal on Computing}, 49(5):STOC18--368--STOC18--394,
  2020.
\newblock \href {https://doi.org/10.1137/18M120275X}
  {\path{doi:10.1137/18M120275X}}.

\bibitem{QuantumAlghoritmicMeasurements}
Dorit Aharonov, Jordan Cotler, and Xiao-Liang Qi.
\newblock Quantum algorithmic measurement.
\newblock {\em Nature Communications}, 13(1), feb 2022.
\newblock \href {https://doi.org/10.1038/s41467-021-27922-0}
  {\path{doi:10.1038/s41467-021-27922-0}}.

\bibitem{AmbainisEmerson2007}
Andris Ambainis and Joseph Emerson.
\newblock Quantum {T}-designs: {T}-wise independence in the quantum world.
\newblock In {\em Proceedings of the Twenty-Second Annual IEEE Conference on
  Computational Complexity}, CCC '07, page 129–140, USA, 2007. IEEE Computer
  Society.
\newblock \href {https://doi.org/10.1109/CCC.2007.26}
  {\path{doi:10.1109/CCC.2007.26}}.

\bibitem{qiskit}
MD~SAJID ANIS et~al.
\newblock Qiskit: An open-source framework for quantum computing, 2021.
\newblock \href {https://doi.org/10.5281/zenodo.2573505}
  {\path{doi:10.5281/zenodo.2573505}}.

\bibitem{Google2019}
Frank Arute et~al.
\newblock Quantum supremacy using a programmable superconducting processor.
\newblock {\em Nature}, 574(7779):505--510, 10 2019.
\newblock \href {https://doi.org/10.1038/s41586-019-1666-5}
  {\path{doi:10.1038/s41586-019-1666-5}}.

\bibitem{bengtsson_zyczkowski_2006}
Ingemar Bengtsson and Karol Zyczkowski.
\newblock {\em Geometry of Quantum States: An Introduction to Quantum
  Entanglement}.
\newblock Cambridge University Press, 2006.
\newblock \href {https://doi.org/10.1017/CBO9780511535048}
  {\path{doi:10.1017/CBO9780511535048}}.

\bibitem{Berger1997}
Bonnie Berger.
\newblock The fourth moment method.
\newblock {\em SIAM Journal on Computing}, 26(4):1188--1207, 1997.
\newblock \href {https://doi.org/10.1137/S0097539792240005}
  {\path{doi:10.1137/S0097539792240005}}.

\bibitem{Boixo2018}
Sergio Boixo, Sergei~V. Isakov, Vadim~N. Smelyanskiy, Ryan Babbush, Nan Ding,
  Zhang Jiang, Michael~J. Bremner, John~M. Martinis, and Hartmut Neven.
\newblock Characterizing quantum supremacy in near-term devices.
\newblock {\em Nature Physics}, 14(6):595--600, 6 2018.
\newblock \href {https://doi.org/10.1038/s41567-018-0124-x}
  {\path{doi:10.1038/s41567-018-0124-x}}.

\bibitem{LocalRandomCircuitsDesigns}
Fernando G.~S.~L. {Brand{\~a}o}, Aram~W. {Harrow}, and Micha{\l} {Horodecki}.
\newblock Local random quantum circuits are approximate polynomial-designs.
\newblock {\em Communications in Mathematical Physics}, 346(2):397--434, 9
  2016.
\newblock \href {https://doi.org/10.1007/s00220-016-2706-8}
  {\path{doi:10.1007/s00220-016-2706-8}}.

\bibitem{ComplexityGrowthModels}
Fernando~G.S.L. Brand\~ao, Wissam Chemissany, Nicholas Hunter-Jones, Richard
  Kueng, and John Preskill.
\newblock Models of quantum complexity growth.
\newblock {\em PRX Quantum}, 2:030316, 7 2021.
\newblock \href {https://doi.org/10.1103/PRXQuantum.2.030316}
  {\path{doi:10.1103/PRXQuantum.2.030316}}.

\bibitem{Chen2021shadow}
Senrui Chen, Wenjun Yu, Pei Zeng, and Steven~T. Flammia.
\newblock Robust shadow estimation.
\newblock {\em PRX Quantum}, 2(3), 9 2021.
\newblock \href {https://doi.org/10.1103/prxquantum.2.030348}
  {\path{doi:10.1103/prxquantum.2.030348}}.

\bibitem{Emerson2005RB}
Joseph Emerson, Robert Alicki, and Karol Życzkowski.
\newblock Scalable noise estimation with random unitary operators.
\newblock {\em Journal of Optics B: Quantum and Semiclassical Optics},
  7(10):S347–S352, 9 2005.
\newblock \href {https://doi.org/10.1088/1464-4266/7/10/021}
  {\path{doi:10.1088/1464-4266/7/10/021}}.

\bibitem{farhi2014qaoa}
Edward Farhi, Jeffrey Goldstone, and Sam Gutmann.
\newblock {A} quantum approximate optimization algorithm, 2014.
\newblock \href {http://arxiv.org/abs/1411.4028} {\path{arXiv:1411.4028}}.

\bibitem{farhi2019quantum}
Edward Farhi and Aram~W Harrow.
\newblock Quantum supremacy through the quantum approximate optimization
  algorithm, 2019.
\newblock \href {http://arxiv.org/abs/1602.07674} {\path{arXiv:1602.07674}}.

\bibitem{flammia2021ACES}
Steven~T. Flammia.
\newblock Averaged circuit eigenvalue sampling, 2021.
\newblock \href {http://arxiv.org/abs/2108.05803} {\path{arXiv:2108.05803}}.

\bibitem{Gambetta2012RB}
Jay~M. Gambetta, A.~D. Córcoles, S.~T. Merkel, B.~R. Johnson, John~A. Smolin,
  Jerry~M. Chow, Colm~A. Ryan, Chad Rigetti, S.~Poletto, Thomas~A. Ohki, and
  et~al.
\newblock Characterization of addressability by simultaneous randomized
  benchmarking.
\newblock {\em Physical Review Letters}, 109(24), 12 2012.
\newblock \href {https://doi.org/10.1103/physrevlett.109.240504}
  {\path{doi:10.1103/physrevlett.109.240504}}.

\bibitem{Garcia-Perez2021learning}
Guillermo García-Pérez, Matteo~A.C. Rossi, Boris Sokolov, Francesco Tacchino,
  Panagiotis~Kl. Barkoutsos, Guglielmo Mazzola, Ivano Tavernelli, and Sabrina
  Maniscalco.
\newblock Learning to measure: Adaptive informationally complete generalized
  measurements for quantum algorithms.
\newblock {\em PRX Quantum}, 2(4), 11 2021.
\newblock \href {https://doi.org/10.1103/prxquantum.2.040342}
  {\path{doi:10.1103/prxquantum.2.040342}}.

\bibitem{Distances2005}
Alexei {Gilchrist}, Nathan~K. {Langford}, and Michael~A. {Nielsen}.
\newblock Distance measures to compare real and ideal quantum processes.
\newblock {\em Phys.\ Rev.\ A}, 71(6):062310, 6 2005.
\newblock \href {https://doi.org/10.1103/PhysRevA.71.062310}
  {\path{doi:10.1103/PhysRevA.71.062310}}.

\bibitem{hadfield2021shadow}
Charles Hadfield.
\newblock Adaptive pauli shadows for energy estimation, 2021.
\newblock \href {http://arxiv.org/abs/2105.12207} {\path{arXiv:2105.12207}}.

\bibitem{hadfield2020shadow}
Charles Hadfield, Sergey Bravyi, Rudy Raymond, and Antonio Mezzacapo.
\newblock Measurements of quantum hamiltonians with locally-biased classical
  shadows.
\newblock {\em Communications in Mathematical Physics}, 391(3):951--967, May
  2022.
\newblock \href {https://doi.org/10.1007/s00220-022-04343-8}
  {\path{doi:10.1007/s00220-022-04343-8}}.

\bibitem{ExplicitDesignsNickJonas2021}
Jonas Haferkamp and Nicholas Hunter-Jones.
\newblock Improved spectral gaps for random quantum circuits: Large local
  dimensions and all-to-all interactions.
\newblock {\em Phys. Rev. A}, 104:022417, 8 2021.
\newblock \href {https://doi.org/10.1103/PhysRevA.104.022417}
  {\path{doi:10.1103/PhysRevA.104.022417}}.

\bibitem{Hansen1990algorithms}
Pierre Hansen and Brigitte Jaumard.
\newblock Algorithms for the maximum satisfiability problem.
\newblock {\em Computing}, 44(4):279--303, 12 1990.
\newblock \href {https://doi.org/10.1007/BF02241270}
  {\path{doi:10.1007/BF02241270}}.

\bibitem{Harrigan2021QAOA}
Matthew~P. Harrigan et~al.
\newblock Quantum approximate optimization of non-planar graph problems on a
  planar superconducting processor.
\newblock {\em Nature Physics}, 17(3):332--336, feb 2021.
\newblock \href {https://doi.org/10.1038/s41567-020-01105-y}
  {\path{doi:10.1038/s41567-020-01105-y}}.

\bibitem{harrow2013church}
Aram~W. Harrow.
\newblock The church of the symmetric subspace, 2013.
\newblock \href {http://arxiv.org/abs/1308.6595} {\path{arXiv:1308.6595}}.

\bibitem{DesignsShallow2018}
Aram~W. Harrow and Saeed Mehraban.
\newblock Approximate unitary t-designs by short random quantum circuits using
  nearest-neighbor and long-range gates.
\newblock {\em Communications in Mathematical Physics}, 401(2):1531--1626, may
  2023.
\newblock \href {https://doi.org/10.1007/s00220-023-04675-z}
  {\path{doi:10.1007/s00220-023-04675-z}}.

\bibitem{helsen2019RB}
Jonas Helsen, Xiao Xue, Lieven M.~K. Vandersypen, and Stephanie Wehner.
\newblock A new class of efficient randomized benchmarking protocols.
\newblock {\em npj Quantum Information}, 5(1):71, Aug 2019.
\newblock \href {https://doi.org/10.1038/s41534-019-0182-7}
  {\path{doi:10.1038/s41534-019-0182-7}}.

\bibitem{DiamondNormQEC2019}
Eric {Huang}, Andrew~C. {Doherty}, and Steven {Flammia}.
\newblock Performance of quantum error correction with coherent errors.
\newblock {\em Phys.\ Rev.\ A}, 99(2):022313, 2 2019.
\newblock \href {https://doi.org/10.1103/PhysRevA.99.022313}
  {\path{doi:10.1103/PhysRevA.99.022313}}.

\bibitem{Huang2020predicting}
Hsin-Yuan Huang, Richard Kueng, and John Preskill.
\newblock Predicting many properties of a quantum system from very few
  measurements.
\newblock {\em Nature Physics}, 16(10):1050–1057, 6 2020.
\newblock \href {https://doi.org/10.1038/s41567-020-0932-7}
  {\path{doi:10.1038/s41567-020-0932-7}}.

\bibitem{QuantumAdvML2021}
Hsin-Yuan {Huang}, Richard {Kueng}, and John {Preskill}.
\newblock Information-theoretic bounds on quantum advantage in machine
  learning.
\newblock {\em Phys.\ Rev.\ Lett.}, 126(19):190505, 5 2021.
\newblock \href {https://doi.org/10.1103/PhysRevLett.126.190505}
  {\path{doi:10.1103/PhysRevLett.126.190505}}.

\bibitem{Jensen1906}
J.~L. W.~V. Jensen.
\newblock Sur les fonctions convexes et les inégalités entre les valeurs
  moyennes.
\newblock {\em Acta Mathematica}, 30(none):175 -- 193, 1906.
\newblock \href {https://doi.org/10.1007/BF02418571}
  {\path{doi:10.1007/BF02418571}}.

\bibitem{qutip1}
J.~R. Johansson, P.~D. Nation, and Franco Nori.
\newblock {QuTiP}: An open-source python framework for the dynamics of open
  quantum systems.
\newblock {\em Computer Physics Communications}, 183(8):1760--1772, 2012.
\newblock \href {https://doi.org/https://doi.org/10.1016/j.cpc.2012.02.021}
  {\path{doi:https://doi.org/10.1016/j.cpc.2012.02.021}}.

\bibitem{qutip2}
J.~R. Johansson, P.~D. Nation, and Franco Nori.
\newblock {QuTiP} {2}: {A} python framework for the dynamics of open quantum
  systems.
\newblock {\em Computer Physics Communications}, 184(4):1234--1240, 2013.
\newblock \href {https://doi.org/https://doi.org/10.1016/j.cpc.2012.11.019}
  {\path{doi:https://doi.org/10.1016/j.cpc.2012.11.019}}.

\bibitem{Kandala2017VQE}
Abhinav Kandala, Antonio Mezzacapo, Kristan Temme, Maika Takita, Markus Brink,
  Jerry~M. Chow, and Jay~M. Gambetta.
\newblock Hardware-efficient variational quantum eigensolver for small
  molecules and quantum magnets.
\newblock {\em Nature}, 549(7671):242--246, sep 2017.
\newblock \href {https://doi.org/10.1038/nature23879}
  {\path{doi:10.1038/nature23879}}.

\bibitem{CliffordOrbitsDisting2016}
Richard {Kueng}, Huangjun {Zhu}, and David {Gross}.
\newblock Distinguishing quantum states using clifford orbits.
\newblock {\em arXiv e-prints}, 9 2016.
\newblock \href {http://arxiv.org/abs/1609.08595} {\path{arXiv:1609.08595}}.

\bibitem{Lundeen2008}
J.~S. Lundeen, A.~Feito, H.~Coldenstrodt-Ronge, K.~L. Pregnell, Ch. Silberhorn,
  T.~C. Ralph, J.~Eisert, M.~B. Plenio, and I.~A. Walmsley.
\newblock Tomography of quantum detectors.
\newblock {\em Nature Physics}, 5:27, 11 2008.
\newblock \href {https://doi.org/10.1038/nphys1133}
  {\path{doi:10.1038/nphys1133}}.

\bibitem{Maciejewski2021}
Filip~B. Maciejewski, Flavio Baccari, Zoltán Zimborás, and Micha\l{}
  Oszmaniec.
\newblock Modeling and mitigation of cross-talk effects in readout noise with
  applications to the quantum approximate optimization algorithm.
\newblock {\em Quantum}, 5:464, 6 2021.
\newblock \href {https://doi.org/10.22331/q-2021-06-01-464}
  {\path{doi:10.22331/q-2021-06-01-464}}.

\bibitem{technicalVERSION}
Filip~B. Maciejewski, Zbigniew Puchała, and Michał Oszmaniec.
\newblock Exploring quantum average-case distances: Proofs, properties, and
  examples.
\newblock {\em IEEE Transactions on Information Theory}, 69(7):4600--4619,
  2023.
\newblock \href {https://doi.org/10.1109/TIT.2023.3250100}
  {\path{doi:10.1109/TIT.2023.3250100}}.

\bibitem{Maciejewski2020}
Filip~B. Maciejewski, Zolt{\'{a}}n Zimbor{\'{a}}s, and Micha{\l{}} Oszmaniec.
\newblock Mitigation of readout noise in near-term quantum devices by classical
  post-processing based on detector tomography.
\newblock {\em {Quantum}}, 4:257, 4 2020.
\newblock \href {https://doi.org/10.22331/q-2020-04-24-257}
  {\path{doi:10.22331/q-2020-04-24-257}}.

\bibitem{Easwar2010RB}
Easwar Magesan, J.~M. Gambetta, and Joseph Emerson.
\newblock Scalable and robust randomized benchmarking of quantum processes.
\newblock {\em Physical Review Letters}, 106(18), 5 2011.
\newblock \href {https://doi.org/10.1103/physrevlett.106.180504}
  {\path{doi:10.1103/physrevlett.106.180504}}.

\bibitem{Magesan2012RB}
Easwar Magesan, Jay~M. Gambetta, B.~R. Johnson, Colm~A. Ryan, Jerry~M. Chow,
  Seth~T. Merkel, Marcus~P. da~Silva, George~A. Keefe, Mary~B. Rothwell,
  Thomas~A. Ohki, and et~al.
\newblock Efficient measurement of quantum gate error by interleaved randomized
  benchmarking.
\newblock {\em Physical Review Letters}, 109(8), 8 2012.
\newblock \href {https://doi.org/10.1103/physrevlett.109.080505}
  {\path{doi:10.1103/physrevlett.109.080505}}.

\bibitem{Barren2018}
Jarrod~R. {McClean}, Sergio {Boixo}, Vadim~N. {Smelyanskiy}, Ryan {Babbush},
  and Hartmut {Neven}.
\newblock Barren plateaus in quantum neural network training landscapes.
\newblock {\em Nature Communications}, 9:4812, 11 2018.
\newblock \href {https://doi.org/10.1038/s41467-018-07090-4}
  {\path{doi:10.1038/s41467-018-07090-4}}.

\bibitem{NavasquesOperational2014}
Miguel Navascu\'es and Sandu Popescu.
\newblock How energy conservation limits our measurements.
\newblock {\em Phys. Rev. Lett.}, 112:140502, 4 2014.
\newblock \href {https://doi.org/10.1103/PhysRevLett.112.140502}
  {\path{doi:10.1103/PhysRevLett.112.140502}}.

\bibitem{Nielsen2010}
Michael~A. Nielsen and Isaac~L. Chuang.
\newblock {\em Quantum Computation and Quantum Information: 10th Anniversary
  Edition}.
\newblock Cambridge University Press, 2010.
\newblock \href {https://doi.org/10.1017/CBO9780511976667}
  {\path{doi:10.1017/CBO9780511976667}}.

\bibitem{Oszmaniec17}
Micha\l{} Oszmaniec, Leonardo Guerini, Peter Wittek, and Antonio Ac\'{\i}n.
\newblock Simulating positive-operator-valued measures with projective
  measurements.
\newblock {\em Phys. Rev. Lett.}, 119:190501, 11 2017.
\newblock \href {https://doi.org/10.1103/PhysRevLett.119.190501}
  {\path{doi:10.1103/PhysRevLett.119.190501}}.

\bibitem{designsNETS}
Micha\l{} Oszmaniec, Adam Sawicki, and Micha\l{} Horodecki.
\newblock Epsilon-nets, unitary designs and random quantum circuits.
\newblock {\em IEEE Transactions on Information Theory}, pages 1--1, 2021.
\newblock \href {https://doi.org/10.1109/TIT.2021.3128110}
  {\path{doi:10.1109/TIT.2021.3128110}}.

\bibitem{Parrish2019VQE}
Robert~M. Parrish, Edward~G. Hohenstein, Peter~L. McMahon, and Todd~J.
  Mart\'{\i}nez.
\newblock Quantum computation of electronic transitions using a variational
  quantum eigensolver.
\newblock {\em Phys. Rev. Lett.}, 122:230401, 6 2019.
\newblock \href {https://doi.org/10.1103/PhysRevLett.122.230401}
  {\path{doi:10.1103/PhysRevLett.122.230401}}.

\bibitem{peruzzo2014vqe}
Alberto Peruzzo, Jarrod McClean, Peter Shadbolt, Man-Hong Yung, Xiao-Qi Zhou,
  Peter~J. Love, Alán Aspuru-Guzik, and Jeremy~L. O’Brien.
\newblock {A} variational eigenvalue solver on a photonic quantum processor.
\newblock {\em Nature Communications}, 5(1), 7 2014.
\newblock \href {https://doi.org/10.1038/ncomms5213}
  {\path{doi:10.1038/ncomms5213}}.

\bibitem{Preskill2018}
John Preskill.
\newblock Quantum {C}omputing in the {NISQ} era and beyond.
\newblock {\em {Quantum}}, 2:79, 8 2018.
\newblock \href {https://doi.org/10.22331/q-2018-08-06-79}
  {\path{doi:10.22331/q-2018-08-06-79}}.

\bibitem{Puchala2018optimal}
Zbigniew Pucha\l{}a, \L{}ukasz Pawela, Aleksandra Krawiec, and Ryszard
  Kukulski.
\newblock Strategies for optimal single-shot discrimination of quantum
  measurements.
\newblock {\em Physical Review A}, 98(4), 10 2018.
\newblock \href {https://doi.org/10.1103/physreva.98.042103}
  {\path{doi:10.1103/physreva.98.042103}}.

\bibitem{watrous2009semidefinite}
John Watrous.
\newblock Semidefinite programs for completely bounded norms.
\newblock {\em Theory of Computing}, 5(11):217--238, 2009.
\newblock \href {https://doi.org/10.4086/toc.2009.v005a011}
  {\path{doi:10.4086/toc.2009.v005a011}}.

\bibitem{chinesesupreme2021}
Qingling Zhu et~al.
\newblock Quantum computational advantage via 60-qubit 24-cycle random circuit
  sampling.
\newblock {\em Science Bulletin}, 67(3):240--245, 2022.
\newblock \href {https://doi.org/10.1016/j.scib.2021.10.017}
  {\path{doi:10.1016/j.scib.2021.10.017}}.

\end{thebibliography}
\onecolumngrid
\appendix

\section{Worst-case quantum distances}\label{app:sec_optimal_distances}

As mentioned in the main text, commonly used distance measures are based on optimal statistical distinguishability of the objects in question.
We have the following statistical interpretations of trace distance $\dtr$ between quantum states \cite{Nielsen2010}, operational distance $\dop$ \cite{NavasquesOperational2014,Puchala2018optimal} between quantum measurements, and the diamond norm distance  $\ddiam$ \cite{Nielsen2010} between quantum channels
\begin{eqnarray}
    \dtr(\rho,\sigma)&= &\max_{\M} \dtv(\p^{\rho,\M},\p^{\sigma,\M})\ , \\
    \dop(\M,\N)&= & \max_{\rho} \dtv(\p^{\rho,\M},\p^{\rho,\N})\ , \\
    \ddiam(\Lambda,\Gamma)& = & \max_{\rho,\ \M} \dtv(\p^{\rho,\Lambda,\M},\p^{\rho,\Gamma,\M})\ .
\end{eqnarray}
For the case of states, the maximization is over POVMs $\M$ used to distinguish them.
We have a dual situation for measurements, the maximization is over input quantum states used to differentiate between one POVM and another. 
Finally, for the case of quantum channels and the diamond norm -- the maximization is over both input states (on a possibly extended system)
and over POVMs applied after a channel is implemented.

\section{Simplified proofs of main Theorems }\label{app:sec_proofs_theorems}
Here we present simplified versions of proofs of Theorems~\ref{th:STATESav}, \ref{th:MEASav}, and \ref{th:CHANNELSav} from the main text. 
We refer the Reader to \cite{technicalVERSION} for detailed calculations.
Since in the main text we omitted dependence on $\delta$ in $\delta$-approximate unitary designs, we consider here proofs only for exact (not approximate) unitary designs.
The functional dependence for approximate designs, as well as proofs for approximate designs, can be found in \cite{technicalVERSION}.

\subsection{Lower and upper bounds on absolute values}
In scenarios we consider, we aim to find bounds on a random variable that is a Total-Variation distance (TVD) between two probability distributions. 
Note that since the expectation value is linear, it suffices to focus attention on a single outcome probability, and then add resulting bounds to obtain bounds on TVD.

Let us thus denote by $X_i = p_i-q_i$ the value of a difference of probabilities of measurement outcome $i$ taken from probability distributions $p$ and $q$ that correspond to two quantum-mechanical protocols.
This is a shorthand notation -- the protocols are described in the main text and correspond to discrimination between two states, measurements, or general channels.
Conveniently, it turns out that for considered scenarios and probability measures (Haar measure and unitary designs), one can find real parameters $a$ such that the following holds.
\begin{lem}\label{lem:lower_bound}(Lower bound on absolute value)
\begin{align}\label{eq:lower_bound}
\   a \ \sqrt{(\expect{} [X_i^2])}  \leq \expect{} |X_i| \ ,
\end{align}
where the value of $a$ depends on whether we discriminate between states, measurements, or channels.
\end{lem}
\begin{proof}
From Lemmas~4, and 5 in \cite{technicalVERSION} it follows that one can find constants $a$ such that
\begin{align}\label{eq:app:lower_constant}
    \expect{} [X_i^4] \leq \frac{1}{\sqrt{a}}\ \left(\expect{}[X_i^2]\right)^2 \ .
\end{align}
We note that Lemma~4 from \cite{technicalVERSION} is Lemma 2 from \cite{CliffordOrbitsDisting2016}, while Lemma~5 from \cite{technicalVERSION} is one of the results in the accompanying technical manuscript \cite{technicalVERSION}.
Recall that Berger's inequality \cite{Berger1997} states that for random variable $Y$ with well-defined 2nd and 4th moments, we have
\begin{align}\label{eq:bergers_inequality}
\frac{(\expect{} [Y^2])^\frac{3}{2}}{(\expect{} [Y^4])^\frac{1}{2} } \leq \expect{} |Y|\ ,
\end{align}
Then the proof follows from combining Eq.~\eqref{eq:app:lower_constant} with Berger's inequality. 
\end{proof}
At the same time, we have that the following holds for any random variable $Y$.
\begin{lem}\label{lem:jensens_inequality}(Upper bound on absolute value)
\begin{align*}\label{eq:jensens_inequality}
   \expect{}[|Y|] = \expect{}[\sqrt{Y^2}] \leq \sqrt{\expect{}[Y^2]} \ .
\end{align*}
\end{lem}
\begin{proof}
    The above is a special case of Jensen's inequality \cite{Jensen1906} which states that for a concave function $f$ we have $\expect{}[f\rbracket{Y}] \leq f\rbracket{\expect{}[Y]}$.
\end{proof}

From the above one can see that to obtain both lower and upper bound on TVD it suffices to calculate the 2nd moment of $|X_i|$.
To do so, the following Lemma will be useful.
\begin{lem}[Ancillary integral for 2nd moment]\label{lem:2nd_moment}
Let $A$ be a Hermitian operator on $(\H_\dim)$ and $\mu$ be a Haar measure.
Then we have
\begin{equation}\label{eq:2MomentSimple}
      \expect{U\sim\mu} \left[\tr(\kb{i}{i} U A U^\dagger)^2\right] = \frac{1}{d(d+1)} \left(\tr(A^2)+\tr(A)^2\right)\ .
\end{equation}
\end{lem}
\begin{proof}
We first write simple manipulation
\begin{align}
    \expect{U\sim\mu} \left[\tr(\kb{i}{i} U A U^\dagger)^2\right] = \expect{U\sim\mu} \left[\tr\left((U^\dagger)^{\ot 2}\kb{i}{i}^{\ot 2} U^{\ot 2} A^{\ot 2} \right)\right] \ .
\end{align}
This allows us to evaluate the RHS using standard techniques of Haar measure integration (see, e.g., \cite[Prop. 6]{harrow2013church}), and obtain that it is proportional to $\Psym{2}$, i.e., projector onto $2$-fold symmetric subspace of $\H_\dim^{\ot 2}$.
Then the proof follows from applying identities $\Psym{2}= \frac{1}{2}\left(\iden + \mathbb{S}\right)$ and $\tr\left(\mathbb{S}A^{\ot 2}\right) = \tr\left(A^2\right)$, where $\mathbb{S}=\sum_{i, j =1}^{d} \ketbra{ij}{ji}$ is a generalized SWAP operator.
\end{proof}

\subsection{Proofs of Theorems~\ref{th:STATESav} and \ref{th:MEASav}}
For states and measurements, the proofs are essentially identical, thus we consider them together.
As stated above, obtaining both bounds reduces to calculating second moments of $|X_i|$, which we will now outline.

Consider discrimination of states $\rho$ and $\sigma$.
We calculate the second moment by applying Lemma~\ref{lem:2nd_moment} to operator $\Delta_i = \rho - \sigma$, which yields
\begin{align}\label{eq:app:states}
\expect{U\sim\mu}[X_i^2]=\expect{U\sim\mu} \tr( U^\dagger\kb{i}{i} U \Delta_i )^2 = \frac{1}{d(d+1)}\tr(\Delta^2)\ = \frac{1}{d(d+1)} ||\rho-\sigma||_{\HS}^2.
\end{align}
Note that the RHS does not depend on index $i$.
The proof concludes by taking a square root of the RHS and summing over $i$.

Consider discrimination of measurements $\M$ and $\N$. In analogy to states, we calculate the 2nd moment by applying Lemma~\ref{lem:2nd_moment} to operator $\tilde{\Delta}_i = M_i-N_i$, and obtain 
\begin{align}
\expect{U\sim\mu}[X_i^2]=\expect{U\sim\mu} \tr( U^\dagger\psi_{0} U \tilde{\Delta}_i )^2 = \frac{1}{d(d+1)}\left(\tr(\tilde{\Delta}_i^2)+\tr(\tilde{\Delta}_i)^2\right)\ .
\end{align}

\subsection{Proof of Theorem~\ref{th:CHANNELSav}}
In the case of states and measurements, there was only a single average (over projective measurements for states and over pure states for measurements).
However, for quantum channels we have both quantum inputs and outputs, thus we need to calculate two averages.
Consider discrimination between two channels $\Lambda$ and $\Gamma$.
Denote $\Delta = \Lambda - \Gamma$.

To proceed, we first apply Theorem~\ref{th:STATESav} to perform averaging over projective measurements after the application of the channel (or, equivalently, averaging over unitaries acting on the output of channels followed by fixed measurement in a standard basis).
In this way, we remove one integral and reduce the problem to finding bounds on the expected value of $\expect{\psi\sim\nu_\pstates} \left\|\Delta[\psi]  \right\|_\HS = \expect{\psi\sim\nu_\pstates} \left[\sqrt{\tr\left(\Delta[\psi]^2 \right)}\right]$.
Using the same line of arguments as before, this quantity can be lower and upper bounded by evaluating  $\expect{\psi\sim\nu_\pstates}  \tr\left(\Delta[\psi]^2\right)$.
This is done by first performing simple manipulation
\begin{align}
    \expect{\psi\sim\nu_\pstates}  \left[\tr\left(\Delta[\psi]^2\right) \right]= \expect{\psi\sim\nu_\pstates} \left[\tr\left(\mathbb{S}\Delta[\psi]^{\ot 2}\right) \right] = \tr\left(\mathbb{S}\expect{\psi\sim\nu_\pstates}\left[\Delta[\psi]^{\otimes 2}\right]\right) .
\end{align}
The last term in the above can then be evaluated using standard techniques of Haar measure integration (see, e.g., \cite[Prop. 6]{harrow2013church}, and recall the proof of Lemma~\ref{lem:2nd_moment}).
The computation yields
\begin{equation}\label{eq:chanAVexplicit}
    \expect{\psi\sim\nu_\pstates} \left[\tr(\Delta[\psi]^2 ) \right]
    =\frac{d^2}{d(d+1)}\left(\tr\left(\Delta\left[\frac{\I}{d}\right]^2\right)   + \tr\left(\S\Delta^{\ot 2}\left[\frac{\S}{d^2}\right] \right)\right)\ .
\end{equation}
Noticing that $ \tr\left(\S \Delta^{\ot 2}\left[\frac{\S}{d^2}\right] \right) = \|\J_{\Delta}\|^2_\HS 
$ concludes the proof.

\section{Proofs of claims in Examples 1-4}\label{app:sec_examples}

As mentioned in the main text, Examples 1-5 follow directly from more general expressions in examples in technical manuscript \cite{technicalVERSION}.
Specifically, the Example~1 follows from Example~9, Examples~2 and 3 follow from Example~10 (in case of Example~3 arguments are slightly more involved, as presented below), while Examples~4 and 5 follow from Example~14.

We now recall statements of Example~9 for Reader's convenience.

\begin{exa}\label{ex:uniform_separable_states_app}[Example~9 from \cite{technicalVERSION}]
Consider state $\psi^{\text{pauli}} = \otimes_{i=1}^{N} \ketbra{\pm r_i}{\pm r_i}$, where $r_i \in \left\{x,y,z\right\}$, i.e., $\ket{\pm r_i}$ is any Pauli eigenstate on qubit $i$ (with eigenvalue $+1$ or $-1$.).
Consider tensor product Pauli channel $\Lambda^{\text{pauli}}=\otimes_{i=1}^N \Lambda_{i}^{\text{pauli}}$, where single-qubit channel is $\Lambda^{\text{pauli}}_{i}(\rho) = \sum_{j=1}\ p^{(i)}_j\sigma_j\rho\sigma_j$ with $j \in \left\{1,x,y,z\right\}$, $\sigma_1=\iden$, and $p^{(i)}_j\geq 0$, $\sum_{j}p^{(i)}_j=1$.
Define $q^{(i)} = p^{(i)}_1+p^{(i)}_{r_i}$, i.e., a probability of applying on qubit $i$ a gate that stabilizes the state of that qubit (namely, either identity or Pauli matrix of which $\ket{\pm r_i}$ is an eigenstate).
Furthermore, assume that for each qubit $i$ we have $q^{(i)}\geq\frac{1}{2}$.
Then we have
\begin{align}\label{eq:states_uniform_example_app}
    \davS(\Lambda^{\text{pauli}}(\psi^{\text{pauli}}), \frac{\iden}{d}) = \frac{1}{2} \sqrt{\Pi_{i=1}^{N}\left(1-2q^{(i)}(1-q^{(i)})\right)-\frac{1}{d}}\ ,
\end{align}
\begin{align*}\label{eq:states_pauli_example_app}
        &\davS(\Lambda^{\text{pauli}}(\psi^{\text{pauli}}), \psi^{\text{pauli}})=  \frac{1}{2}\sqrt{1-2\Pi_{i=1}^{N}q^{(i)}+\Pi_{i=1}^{N}(1-2q^{(i)}(1-q^{(i)}))}\ , \numberthis
\end{align*}
\end{exa}

We start by defining function $f^{(i)}=q^{(i)}(1-q^{(i)})$, as well as average noise properties  $q^{av}=\frac{1}{N}\sum_{i=1}^{N}q^{(i)}$ and $f^{av}=\frac{1}{N}\sum_{i=1}^{N}f^{(i)}$.
We then bound Eq.~\eqref{eq:states_uniform_example_app} from above as
\begin{align}
    \sqrt{\Pi_{i=1}^{N}\left(1-2f^{(i)}\right)-\frac{1}{d}} \leq \sqrt{\Pi_{i=1}^{N}\left(1-2f^{(i)}\right)} \ ,
\end{align}
and continue with bounding (positive) expression inside square root as
\begin{align}
    \Pi_{i=1}^{N}\left(1-2f^{(i)}\right) = \left(\sqrt[N]{\Pi_{i=1}^{N}\left(1-2f^{(i)}\right)}\right)^N \leq \left(\frac{\sum_{i=1}^{N}(1-2f^{(i)})}{N}\right)^N = \left(1-2f^{av}\right)^N \leq \exp(-2f^{av}N) \ ,
\end{align}
where in first inequality we used inequality between geometric and arithmetic means together with a fact that $x^N \geq y^N$ for $x>y>0$.
In second inequality we used that for $0\leq x\leq 1$ and $N\geq 1$, we have $(1-x)^N\leq \exp\left(-xN\right)$.
Note that each term $2f^{(i)}$ lies in interval $2f^{(i)} \in \sbracket{0,\frac{1}{2}}$.
Combining everything we obtain
\begin{align}
     \davS(\Lambda^{\text{pauli}}(\psi^{\text{pauli}}), \frac{\iden}{d}) \leq \frac{1}{2}\exp(-f^{av}N) \ ,
\end{align}
which concludes the proof of first bound.

To bound Eq.~\eqref{eq:states_pauli_example_app} from below, we start by again employing inequality between geometric and arithmetic mean, namely
\begin{align}
    1-2\Pi_{i=1}^{N}q^{(i)} 
    = 1-2\left(\sqrt[N]{\Pi_{i=1}^{N}q^{(i)}}\right)^N \geq 1-2\left(\frac{\sum_{i=1}q^{(i)}}{N}\right)^N
    = 1-2\left(q^{av}\right)^{N} \ ,
\end{align}
which after combining with Eq.~\eqref{eq:states_pauli_example_app} yields
\begin{align}
        &\davS(\Lambda^{\text{pauli}}(\psi^{\text{pauli}}), \psi^{\text{pauli}}) \geq  \frac{1}{2}\sqrt{1-2\left(q^{av}\right)^N+\Pi_{i=1}^{N}(1-2q^{(i)}(1-q^{(i)}))}\ \geq \frac{1}{2}\sqrt{1-2\left(q^{av}\right)^N} \ .
\end{align}
The above bound is valid provided that argument is still contained in the domain of square root, i.e., we need to impose
\begin{align}
    1-2\left(q^{av}\right)^N\geq 0 \implies  q^{av} \leq \sqrt[N]{\frac{1}{2}} \ . 
\end{align}
Note that $\sqrt[N]{\frac{1}{2}}\xrightarrow{N\rightarrow \infty} 1$, and since $q^{av}$ is by definition lower than 1, the bound becomes less restrictive for higher system sizes.
For small systems it is valid only for high noise (small $q^{av}$), but in such cases one can simply use the exact expressions from Eqs.~\eqref{eq:states_uniform_example_app} and \eqref{eq:states_pauli_example_app}. 

The exactly same reasoning is applied for Examples 2 and 4, for which all expressions have almost the same functional forms (see \cite{technicalVERSION}).
We now consider bound from Example~3 from the main text, for which the first part of the proof is slightly more involved due to more general noise model considered.
\begin{exa}[Example~3 from the main text]\label{ex:app:meas}
Let $\P=(\kb{\x}{\x})_{\x\in\lbrace0,1\rbrace^N}$ be a computational basis measurement on $N$ qubit system. Let $\M=(M_\x)_{\x\in\lbrace0,1\rbrace^N}$ be a POVM specified by effects $M_\x = \Lambda_1^\dagger(\kb{x_1}{x_1})\otimes \ldots \ot  \Lambda_N^\dagger(\kb{x_N}{x_N})$, where $\Lambda_i$ are quantum channels affecting $i$'th qubit, and $\Lambda^\dagger_i$ is the conjugate of $\Lambda_i$. 
Define classical success probability as $p^{(i)}(k|k) = \tr{\left(\Lambda_{i}^{\dag}\left(\ketbra{x_i}{x_i}\right) \ketbra{x_i}{x_i}\right)}$ and corresponding average  $q_{av}^{(i)} = \frac{p^{i}(0|0)+p^{(i)}(1|1)}{2}$.
Let $q^{av} \coloneqq \frac{1}{N}\sum_{i=1}^{N} q_{av}^{(i)}$.
Assume $q^{(i)}_{av}\geq \frac{1}{2}$ for each qubit $i$ and that $q^{av}\leq \sqrt[N]{\frac{1}{2}}$.
Then we have 
\begin{align}
    \dav^\m(\M,\P) > \frac{1}{2}\sqrt{1- 2(q^{av})^N} \ .
\end{align}
\end{exa}
To prove the above, first one applies maximally-dephasing channel to both measurements and uses data-processing inequality for average-case distance to bound the distance from below by the diagonal part of the POVM $\M$. 
Specifically, define dephased POVM $\Phi_{\mathrm{dep}}(\M)$ via its effects $\Phi_{\mathrm{dep}}(\M)_i=\Phi_{\mathrm{dep}}(M_i)$, where maximally dephasing channel acts on any operator $A$ as $\Phi_{\mathrm{dep}}(A)=\mathrm{diag}(A)$, with $\mathrm{diag}(A)$ denoting diagonal part of $A$.
Note that for compuational basis measurement $\P$ we have $\Phi_{\mathrm{dep}}(\P)=\P$.
Thus we have 
\begin{align}
    \davM(\Phi_{\mathrm{dep}}(\M),\Phi_{\mathrm{dep}}(\P)) \geq \davM(\Phi_{\mathrm{dep}}(\M),\P) \ .
\end{align}
The above allows to treat noise as classical and look only on assignment infidelities for classical states (i.e., error probabilites when measured states are computational-basis states).
Note that, importantly, maximally dephasing channel does not change the product structure of $\M$.
Thus we can treat this dephased POVM
$\Phi_{\mathrm{dep}}(\M)$ 
as related to computational basis measurement via some tensor product stochastic map $\clc=\bigotimes_{i=1}^N \clc^{(i)}$, where $\clc^{(i)}$ acts on $i$th qubit and is specified by two success probabilities $p^{(i)}(0|0)$ and $p^{(i)}(1|1)$ (see, for example, Ref.~\cite{Maciejewski2020} for more details on stochastic readout noise).
Thus we have
\begin{align}
    \davM(\M,\P) \geq \davM(\clc\P,\P) \ ,
\end{align}
where $\clc\P$ is a POVM with $i$th effect given by $(\clc\P)_{i}=\sum_{i}\clc_{ij}\ketbra{j}{j}$ and stochastic map $\clc$ is defined via diagonal elements of original POVM $\M$ (as in discussion above).

Now one applies Lemma~28 from technical version of the work \cite{technicalVERSION} that lower bounds the distance via symmetrized version of $\clc$, where now both error probabilities are the same and equal to $q_{av}^{(i)} = \frac{p^{(i)}(0|0)+p^{(i)}(1|1)}{2}$ (note that this is equivalent to Pauli bitflip channel applied with probability $q_{av}^{(i)}$).
Denote such symmetrized version of $\clc$ as $\clc^{\mathrm{sym}}$.
This gives 
\begin{align}
     \davM(\clc\P,\P) \geq \davM(\clc^{\mathrm{sym}}\P,\P).
\end{align}
Therefore we reduced the lower bound to scenario considered in Example~2 from the main text, for which the bound was proved above.

\section{Details on numerical simulations}
In the main text, we presented numerical results of calculating mean Total-Variation distances over ensembles of random unitaries.
Here we describe how those ensembles were constructed.
In each case, the $p$-layer circuit can be written as
\begin{align}
    U_{p} = \prod_{j=1}^{p}U_{\mathrm{rot},j}\ U_{\mathrm{ent},j} \ .
\end{align}
where $U_{\mathrm{rot},j}$ is a "rotation block" and $U_{\mathrm{ent}}$ is an "entangling block".
Exact form of the evolution, as well as the initial state depend on the ensemble.
We consider two such ensembles:
\begin{enumerate}
    \item Circuits that originate from QAOA instance for fixed Hamiltonian $\mathrm{H}_{\text{2SAT}}$ encoding fixed (random) instance of random MAX-2-SAT problem \cite{Hansen1990algorithms}.
In this case, the initial state is of the form $\ket{+}^{\otimes N}$ with $\ket{+}=\frac{1}{\sqrt{2}}(\ket{0}+\ket{1})$, while 
unitary evolution is given by $U_{\mathrm{rot,j}}\coloneqq U_{\alpha_j} = \exp\left(-i\alpha_j \sum_{k=1}^{N}\sigma_{x}^{(k)}\right)$, and $U_{\mathrm{ent},j}\coloneqq U_{\beta_j}= \exp\left(-i\beta_j\mathrm{H}_{\text{2SAT}}\right)$, with $\sigma_{x}^{(k)}$ being X gate on $k$th qubit.
For each $j$, $\alpha_j$ and $\beta_j$ are $N$-dimensional vectors of parameters chosen randomly from range $\sbracket{-\pi,\pi}$.
\item Circuits of a form of generic Hamiltonian-independent VQE ansatz with initial state being $\ket{0}^{\otimes N}$.
We choose the rotation block to be of the form $U_{\mathrm{rot,j}} \coloneqq U_{\alpha_j} = \bigotimes_{k=1}^{N}\exp\left(-i \alpha_{2j}\  \sigma_{Z}^{(k)}\right)\circ\exp\left(-i \alpha_{2j+1}\  \sigma_{Y}^{(k)}\right)$, where $\sigma_{Y},\sigma_{Z}$ are $Y$ and $Z$ gates.
The entangling block is $U_{\mathrm{ent},j} = U_{\mathrm{ent}} \coloneqq \prod_{k=1}^{N-1}\text{CX}_{k,k+1}$ with $\text{CX}_{k,l}$ denoting $\text{CX}$ gate between qubits $k$ and $l$. 
For each $j$, $\alpha_j$ is a $2N$-dimensional vector of parameters chosen randomly from range $\sbracket{-\pi,\pi}$.
\end{enumerate}

\section{Additional numerical results}
Here we provide some additional plots with numerical results.

\begin{figure}[t!]
\begin{center}
\captionsetup[subfigure]{format=default,singlelinecheck=on,justification=RaggedRight}
\subfloat[\label{fig:app:scaling_with_vqe_optimized_states_ideal}Quantum states, distance to ideal distribution]
        {\includegraphics[width=0.475\textwidth]{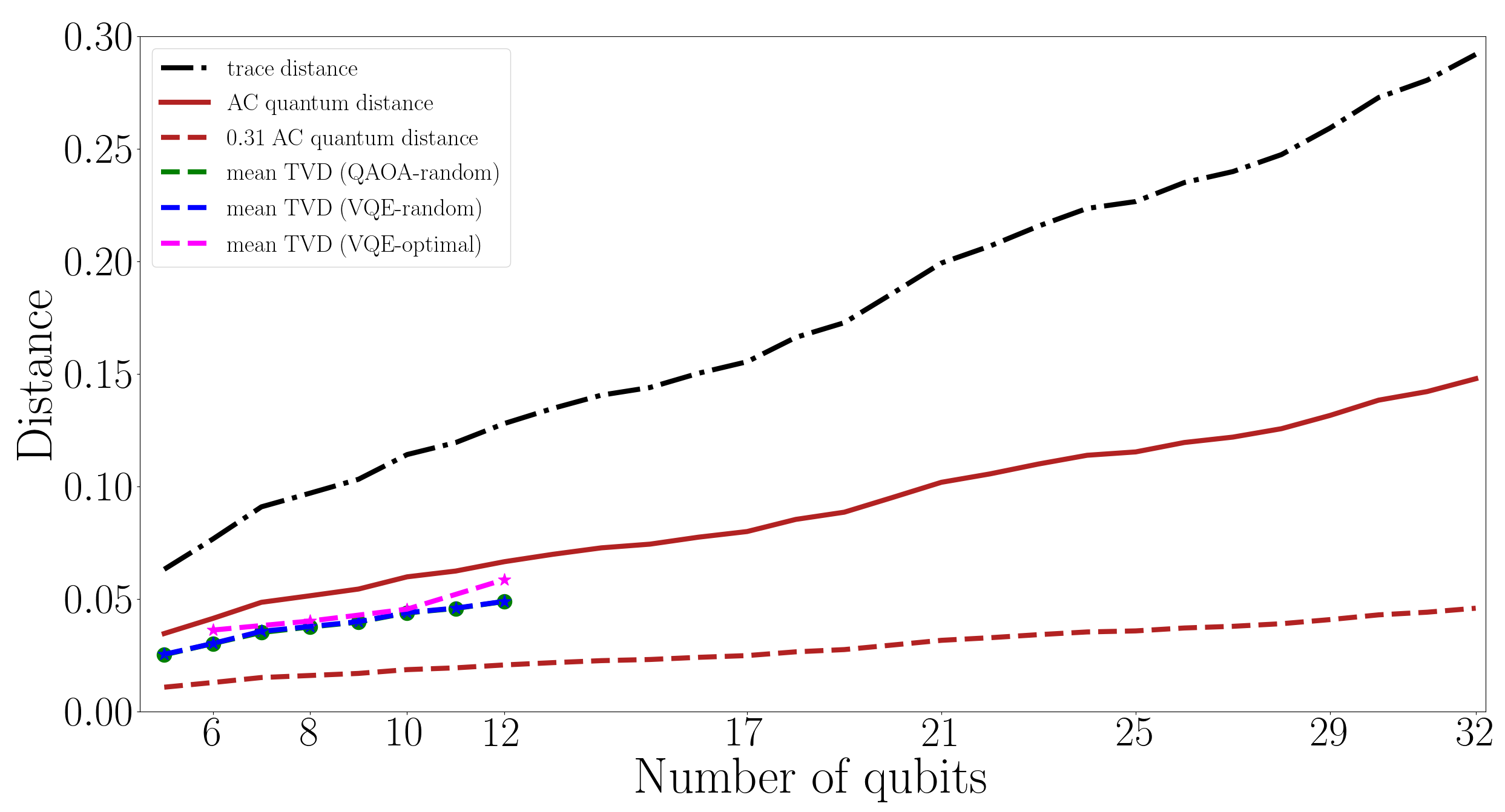}}
   \subfloat[\label{fig:app:scaling_with_vqe_optimized_states_uniform}Quantum states, distance to uniform distribution]
        {\includegraphics[width=0.475\textwidth]{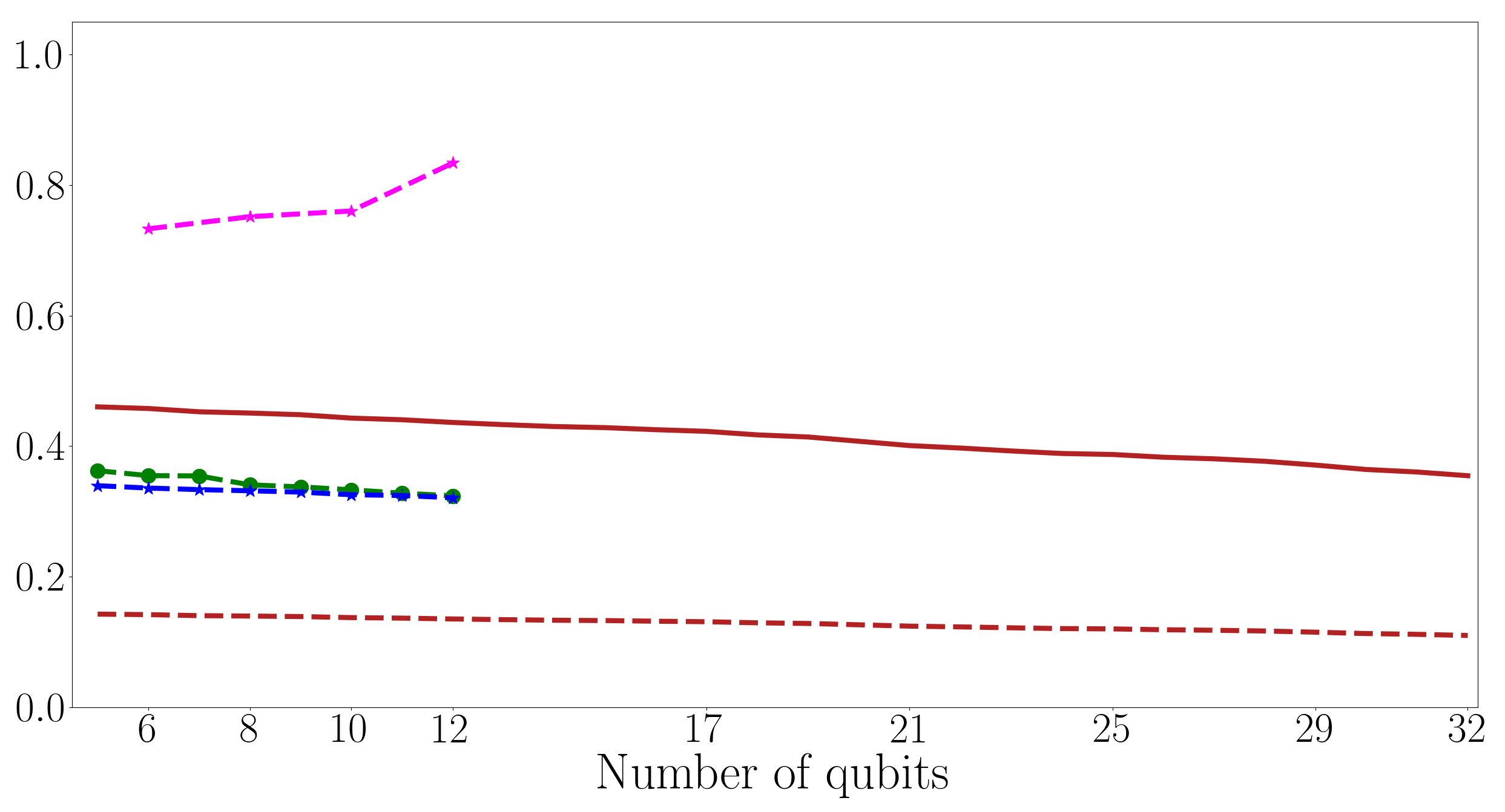}} 
        \\
\subfloat[\label{fig:app:scaling_with_vqe_optimized_measurements_ideal}Quantum measurements, distance to ideal distribution]
    {\includegraphics[width=0.475\textwidth]{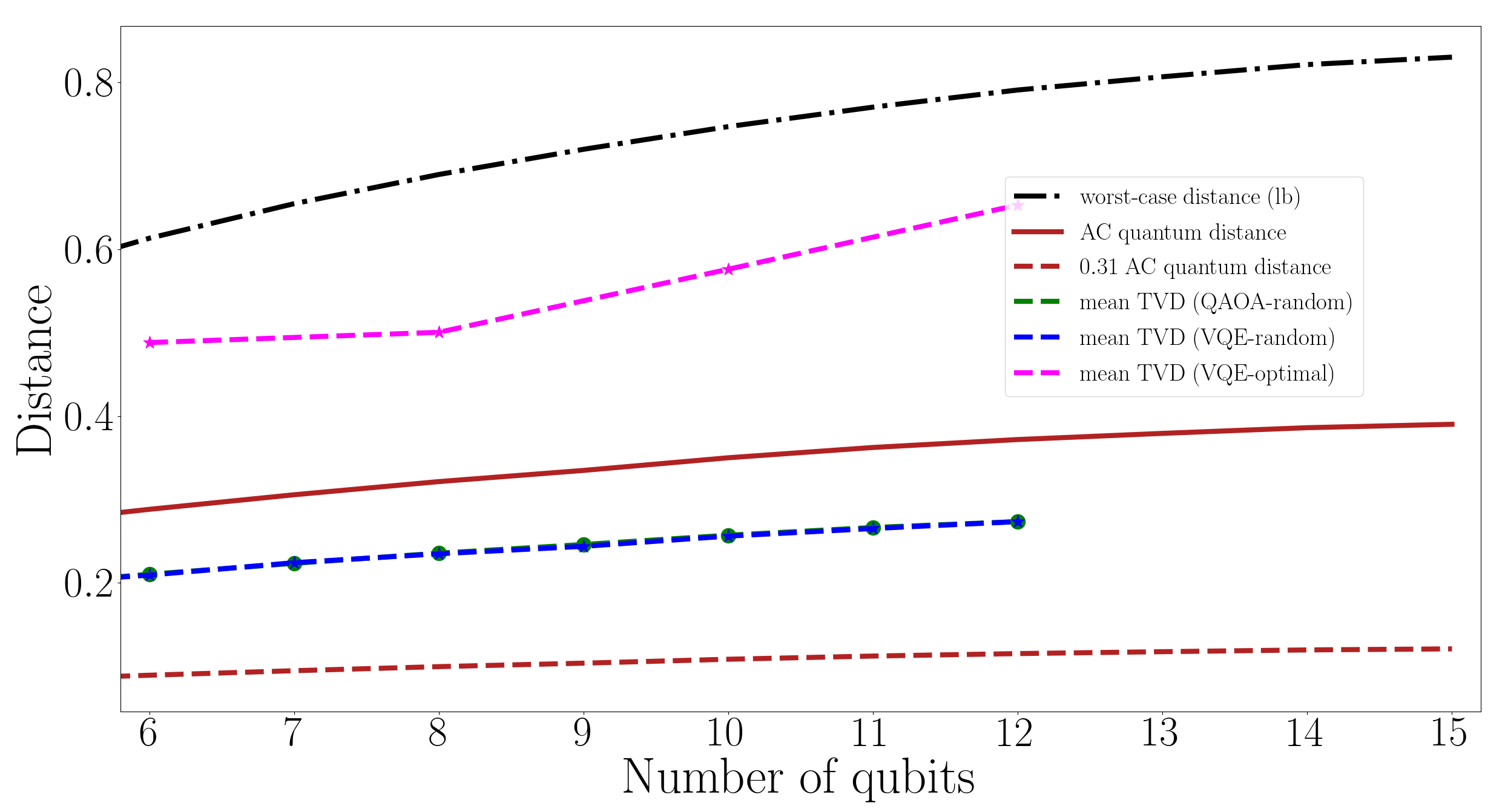}}
\subfloat[\label{fig:app:scaling_with_vqe_optimized_channels_ideal}Quantum channels, distance to ideal distribution]
        {\includegraphics[width=0.475\textwidth]{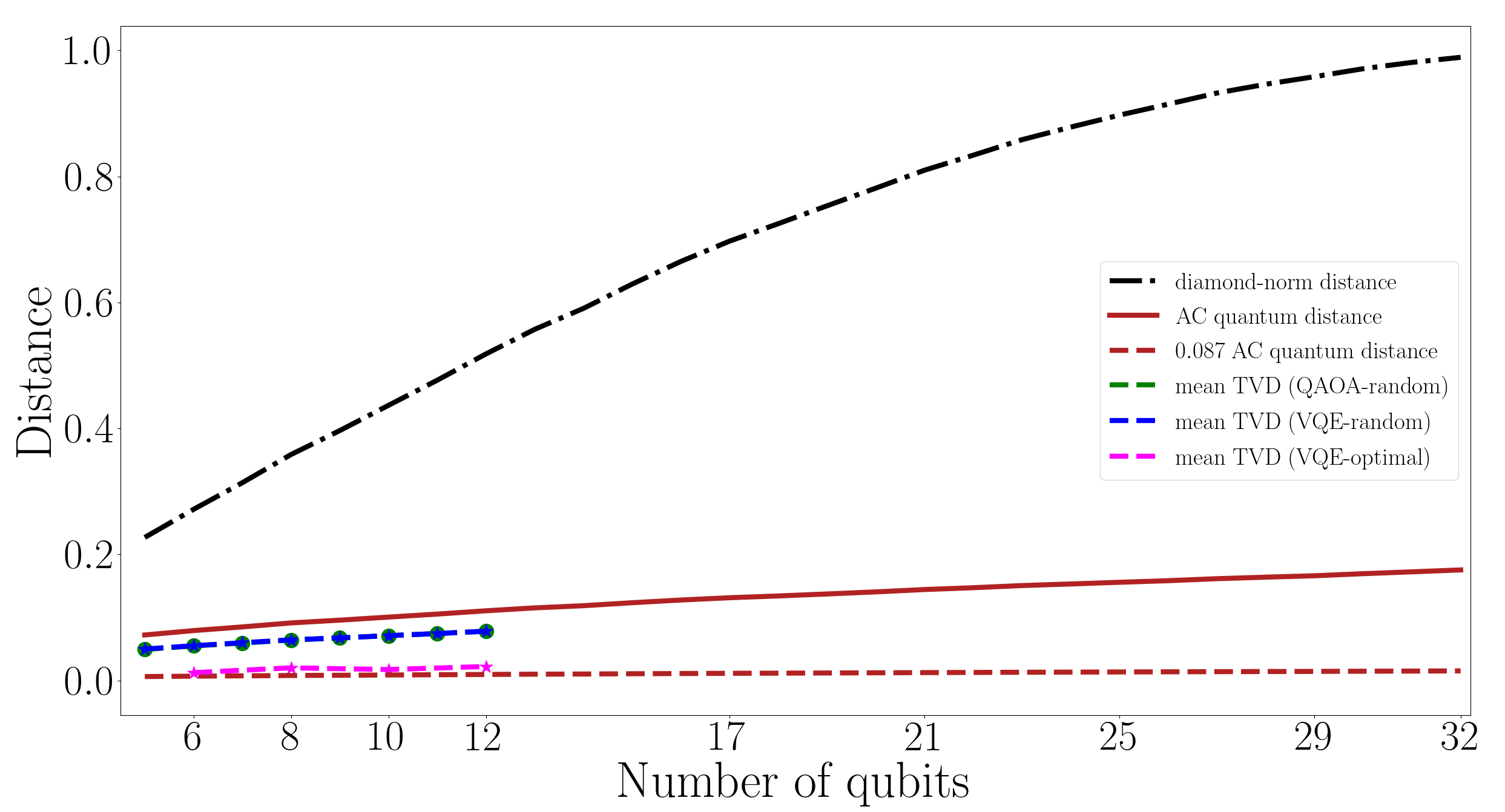}}
    \caption{\label{fig:app:scaling_with_vqe_optimized}
Results of numerical studies for comparison between worst-case distance, average-case quantum distance and numerically calculated mean TVD.
The plot is exactly the same as Fig~\ref{fig:numerics_scaling} in the main text, but with additional ensemble of unitaries considered (see text description).
}
 \end{center}
\end{figure}

In Fig~\ref{fig:app:scaling_with_vqe_optimized} we present the same plots as for Fig~\ref{fig:numerics_scaling} in the main text, but with additional, third ensemble of unitaries considered (see previous section for description of two ensembles used in the main text).

\begin{enumerate}
    \item[3.] The third ensemble is similar to the second VQE-like (see previous section), but now rotation block contains only $Y$ rotations. 
    Furthermore, the angles are \emph{not random}, but they are chosen from a fixed set of parameters that come from solutions of variational optimization.
    In other words, each used unitary corresponds to a circuit that was found to be optimal in a VQE optimization (as opposed to uniformly random angles taken for both previous ensembles).
    We use datasets from Ref.~\cite{Garcia-Perez2021learning} where authors developed an adaptive measurement scheme that improves performance of VQE.
\end{enumerate}

Ensemble of type 3, due to limited computational resources, consist of only $7-12$ unitaries (recall that generating each unitary requires performing full VQE optimization).
This implies that this ensemble \emph{does not} form even unitary $2$-design.
It is nevertheless still interesting to investigate its behaviour, since those are circuits of particular practical importance.

From Fig.~\ref{fig:app:scaling_with_vqe_optimized} we see that for distances between ideal and noisy distributions, the results are qualitatively similar to random ensembles in case of states and channels, but significantly different for quantum measurements.
Recall that POVMs used to generate plot  \ref{fig:app:scaling_with_vqe_optimized_measurements_ideal} are results of detector tomography of actual quantum device from IBM. 
In this case, the noise affects results so much, that empirical TVDs are closer to worst-case than to average-case bounds.
In case of distance between noisy and uniform distribution for states (Fig.~\ref{fig:app:scaling_with_vqe_optimized_states_uniform}) we also observe that average-case distances do not capture well the behaviour of the distributions for unitaries obtained in VQE optimization.

\begin{figure}[!h]
\begin{center}
\captionsetup[subfigure]{format=default,singlelinecheck=on,justification=RaggedRight}
\subfloat[\label{fig:app:histograms_states_ideal}Quantum states, distance to ideal distribution]
        {\includegraphics[width=0.475\textwidth]{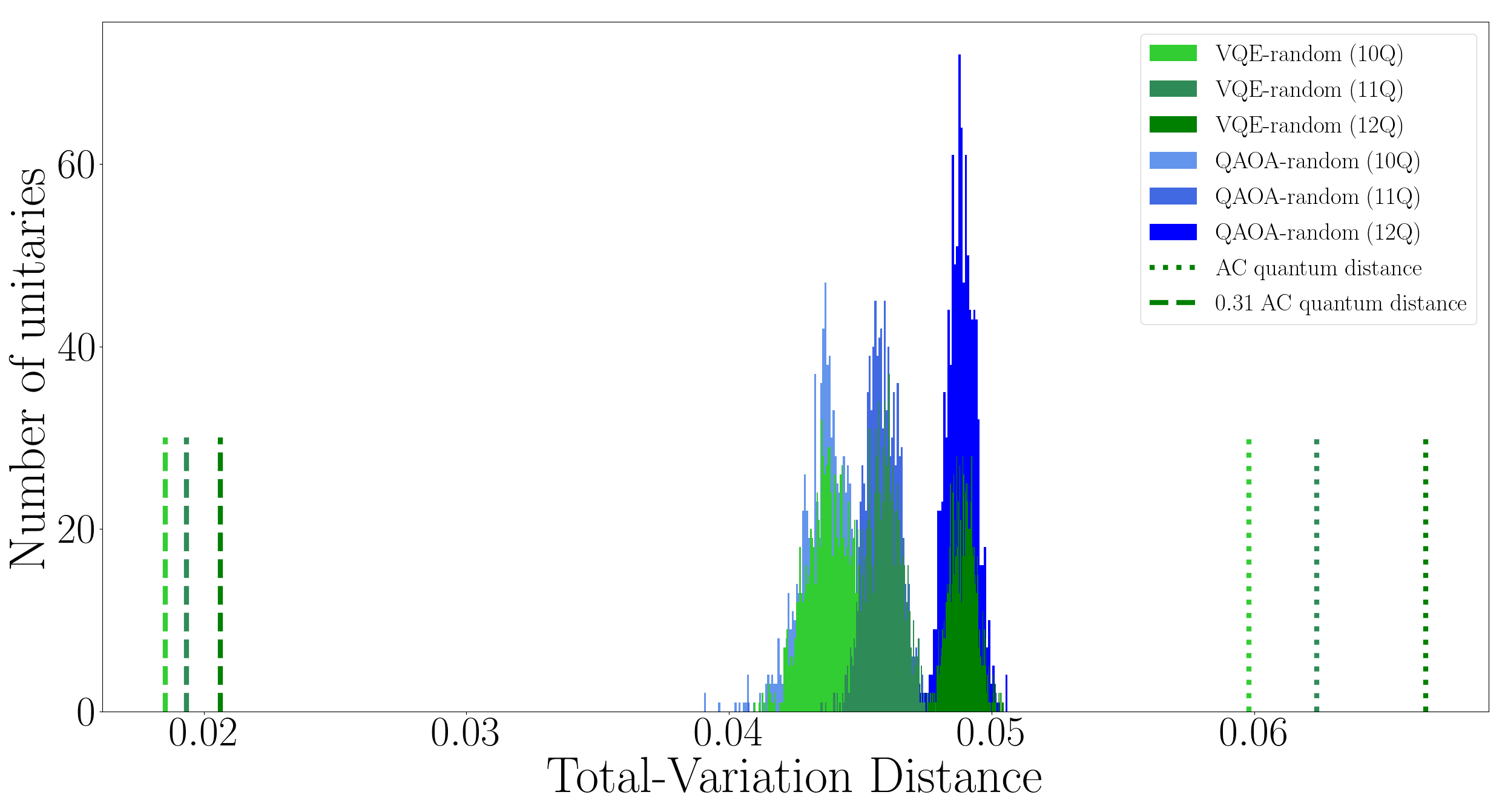}}
\subfloat[\label{fig:app:histograms_measurements_ideal}Quantum measurements, distance to ideal distribution]
    {\includegraphics[width=0.475\textwidth]{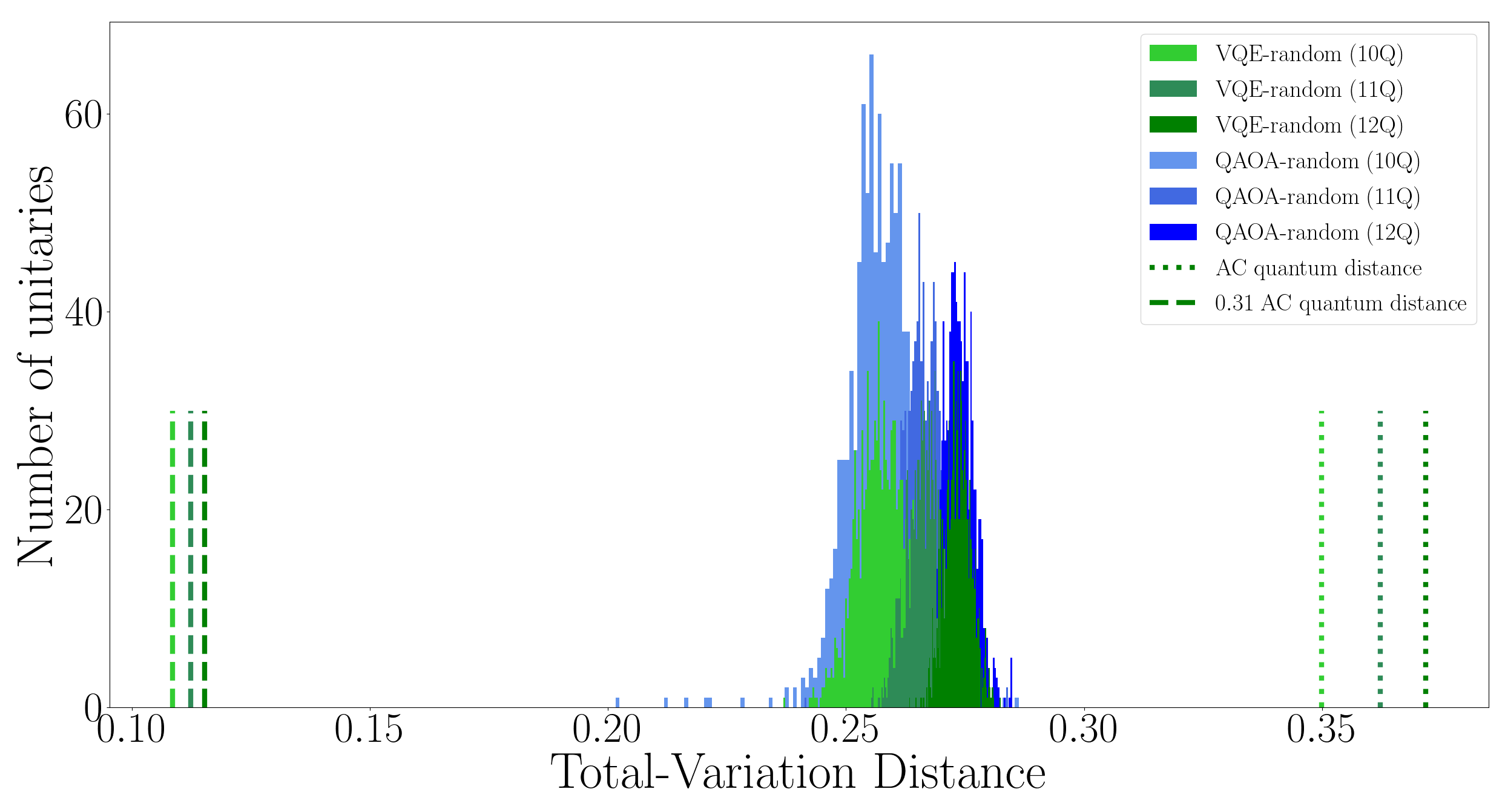}}
\\
\subfloat[\label{fig:app:histograms_channels_ideal}Quantum channels, distance to ideal distribution]
        {\includegraphics[width=0.475\textwidth]{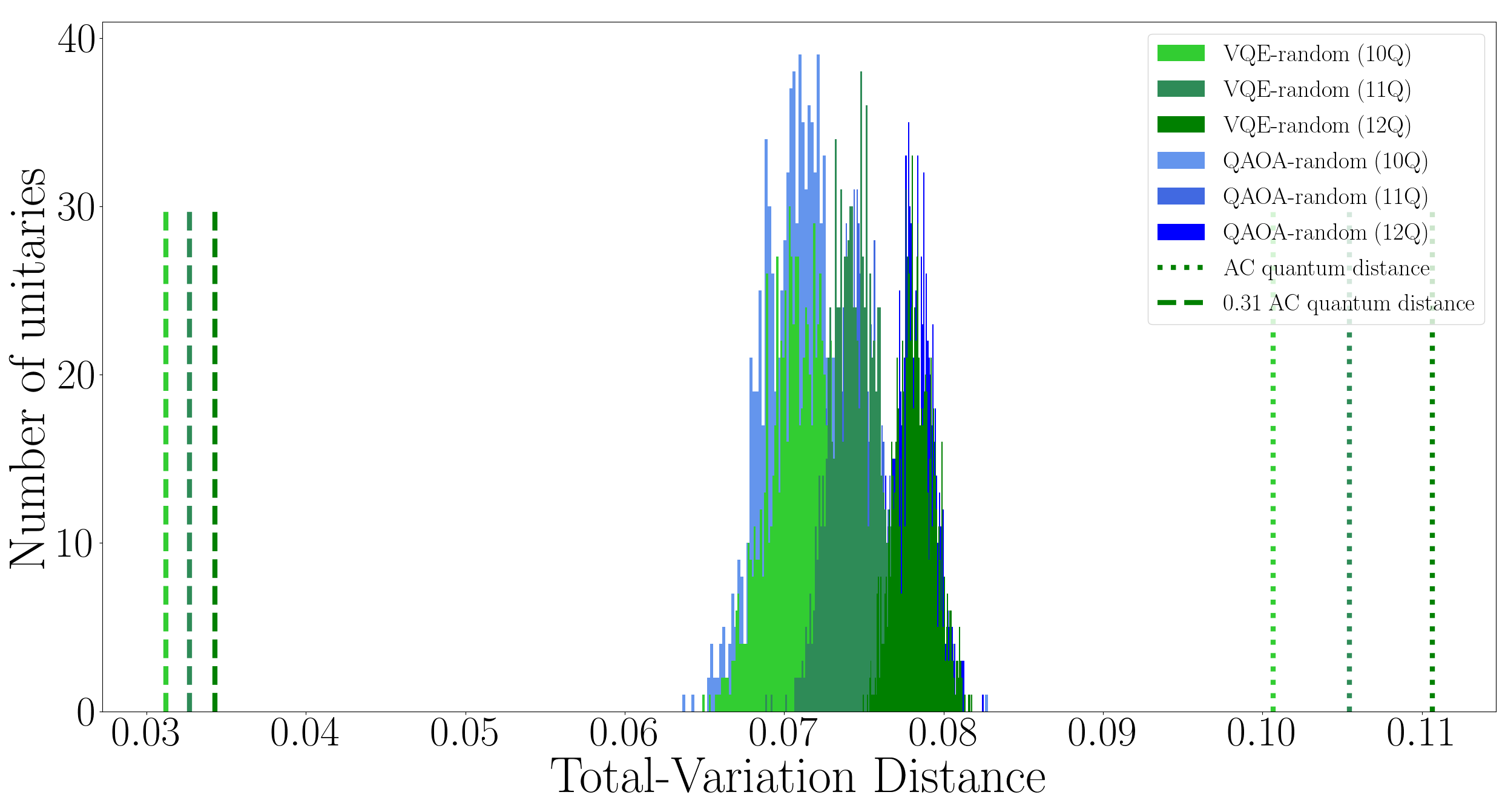}}
    \caption{\label{fig:app:histograms}
    Histograms of TVDs obtained for random ensembles considered in numerical simulations corresponding to Fig~\ref{fig:numerics_scaling} in the main text.
    Different shades of a given color (blue or green) correspond to different system sizes for a given ensemble (QAOA or VQE).
    Bounds from average-case distances are indicated via dashed lines and for each dimension are the same for both ensembles (they depend only on quantum objects in question, not on the choice of random ensemble).
    }
 \end{center}
\end{figure}

In Fig~\ref{fig:app:histograms} we present histograms of TVDs over random unitaries.
The data-points correspond to simulations presented in Fig.~\ref{fig:numerics_scaling} in the main text.
The plots show how the TVDs concentrate for small system sizes and demonstrate that all random points lied well within bounds provided by average-case distances.

\end{document}